\documentclass[sn-mathphys-num]{sn-jnl}% referee option is meant for double line spacing

\usepackage{graphicx}%
\usepackage{multirow}%
\usepackage{amsmath,amssymb,amsfonts}%
\usepackage{amsthm}%
\usepackage{mathrsfs}%
\usepackage[title]{appendix}%
\usepackage{xcolor}%
\usepackage{textcomp}%
\usepackage{manyfoot}%
\usepackage{booktabs}%
\usepackage{algorithm}%
\usepackage{algorithmicx}%
\usepackage{algpseudocode}%
\usepackage{listings}%
\usepackage{placeins}%
\theoremstyle{thmstyleone}%
\theoremstyle{thmstyletwo}%

\theoremstyle{thmstylethree}%

\newcommand{\REVISION}[1]{#1}% (plain for arXiv; red highlighting only in the referee version)
\begin{document}

\title[Article Title]{Quantum Relaxometry Under Continuous Wave Excitation}


\author*[1]{\fnm{Vladimir} \sur{Verkhovlyuk}}\email{vladimir.verkhovlyuk@wigner.hun-ren.hu}

\author[1,2]{\fnm{Chayma} \sur{Bouchair}}%\email{iiauthor@gmail.com}

\author[3]{\fnm{Oleg A.} \sur{Anisimov}}%\email{iiiauthor@gmail.com}

\author*[1,4]{\fnm{Anton} \sur{Pershin}}\email{pershin.anton@wigner.hun-ren.hu}

\author*[1,4,5]{\fnm{Adam} \sur{Gali}}\email{gali.adam@wigner.hun-ren.hu}

\affil[1]{\orgname{HUN-REN Wigner Research Centre for Physics}, \orgaddress{\city{Budapest},  \country{Hungary}}}

\affil[2]{\orgdiv{Gy\"orgy Hevesy Doctoral School, Institute of Chemistry}, \orgname{ELTE E\"otv\"os Lor\'and University}, \orgaddress{\city{Budapest}, \country{Hungary}}}

\affil[3]{\orgdiv{Department}, \orgname{Voevodsky Institute of Chemical Kinetics and Combustion SB RAS}, \orgaddress{\city{Novosibirsk}, \country{Russia}}}

\affil[4]{\orgname{Budapest University of Technology and Economics}, \orgaddress{\city{Budapest},  \country{Hungary}}}

\affil[5]{\orgname{MTA-WFK Lend\"ulet “Momentum” Semiconductor Nanostructures Research Group}, \orgaddress{\city{Budapest}, \country{Hungary}}}

\abstract{
%Quantum relaxometry is among the most successful applications of nitrogen-vacancy centers in diamond and solid-state spin qubits at large, enabling the ultrasensitive detection of magnetic noises and paramagnetic species through measurements of the spin-lattice relaxation time $T_1$. However, conventional pulsed protocols can efficiently probe $T_1$ only within a narrow temporal window, greatly limiting the scope of the technique. Here we introduce a continuous-wave approach to quantum relaxometry that operates in the frequency domain rather than the time domain. By measuring the frequency response of the optically detected magnetic resonance signal under modulated microwave excitation, we extract $T_1$ from the characteristic response time of the spin system. This method enables highly efficient measurements of $T_1$ spanning more than three orders of magnitude, at least from 60~$\mu$s to 200~ms in our experiments with broad region of temperature and inhomogeneity of the spin ensemble. 
%We demonstrate its applicability across different NV platforms including bulk diamond and nanodiamonds, measuring $T_1 \sim 200$~ms in bulk diamond at low temperatures and exceeding 1~ms in nanodiamonds at room temperature. 
Quantum relaxometry is one of the most successful applications of nitrogen-vacancy (NV) centers in diamond and, more broadly, solid-state spin qubits, enabling ultrasensitive detection of magnetic noise and paramagnetic species via measurements of the spin–lattice relaxation time \(T_1\). Conventional pulsed protocols, however, probe \(T_1\) efficiently only over a limited temporal range, which restricts the scope and throughput of the technique. Here we introduce a continuous-wave quantum relaxometry protocol that operates in the frequency domain. By measuring the frequency response of the optically detected magnetic resonance signal under low-frequency microwave amplitude modulation, we extract \(T_1\) from the characteristic response time of the spin system. The method enables efficient \(T_1\) measurements spanning more than three orders of magnitude---\REVISION{directly demonstrated} from 60~\(\mu\)s to 200~ms in our experiments---across a broad temperature range and under substantial ensemble inhomogeneity. 
We further show that this protocol enables quantitative relaxometry-based sensing in nanodiamonds, achieving a \REVISION{substantial speed-up over the pulsed methods} and offering a practical approach to optimizing nanodiamond size for enhanced sensitivity.}

%Using this framework, we further detect high-spin manganese ions in aqueous media at micromolar concentrations and optimise nanodiamond size for enhanced sensitivity to paramagnetic species.}

%\keywords{keyword1, Keyword2, Keyword3, Keyword4}



\maketitle

\section*{Main}

Quantum relaxometry exploits quantum spins as local probes, leveraging the sensitivity of the spin–lattice relaxation time ($T_1$) to magnetic and electric noise, temperature, and the surrounding chemical environment \cite{Rondin_2014,fujisaku2019ph,freire2023role, mzyk2022relaxometry,ariyaratne2018nanoscale}. Negatively charged nitrogen–vacancy (NV$^-$) centres in diamond provide a particularly versatile platform for this purpose: their spin states can be optically polarised and read out at room temperature, coherently manipulated with microwaves, and implemented in both scanning-probe and wide-field imaging configurations \cite{DOHERTY20131,Rondin_2014}. Conventionally, $T_1$ is measured using pulsed time-domain protocols in which a laser pulse initialises the spin into the $|m_s=0\rangle$ state, an optional microwave $\pi$-pulse transfers population to $|m_s=\pm1\rangle$, the system evolves in the dark for a variable delay $\tau$, and a final laser pulse reads out the remaining population \cite{PhysRevLett.108.197601,PhysRevB.108.075411,mzyk2022relaxometry}. The resulting fluorescence decay as a function of $\tau$ yields an exponential relaxation characterised by $T_1$. In practice, accurate determination of $T_1$ also involves complementary measurements with and without the $\pi$-pulse to account for background fluorescence contributions, and requires that each experimental cycle be preceded by a delay long compared with $T_1$ to ensure full relaxation of the spin population \cite{DeichmannGracien2018}.

Extending NV relaxometry to the nanoscale is highly attractive, as it enables spatially resolved access to local noise sources and environmental dynamics \cite{mzyk2022relaxometry,Feng31122022}. However, at these length scales, the conventional pulsed time-domain approach becomes increasingly challenging because of intrinsic limitations in optical collection efficiency, microwave control fidelity, and spin coherence. Measurements on nanodiamonds or near-surface NV centres require fast optical and microwave switching, precise pulse sequencing, and time-resolved detection, all of which become technically demanding as device dimensions shrink and signal levels decrease. Moreover, extracted relaxation times are highly susceptible to errors arising from imperfect pulse timing, residual illumination during the nominal dark interval, NV charge-state dynamics, and microwave pulse imperfections. 
In nanodiamond ensembles, the random orientation of NV axes further complicates the application of well-defined $\pi$-pulses and, in many cases, renders them impractical. \REVISION{In fact, many practical applications measure the relaxation dynamics without a $\pi$-pulse; however, in this case the signal is contaminated by NV$^0$ background and possibly other processes (such as charge-state conversion, repolarization, etc.).}
Moreover, even in bulk time-domain measurements can become prohibitively slow for NV centres with long $T_1$, for instance at low temperature, due to the full spin relaxation requirement \cite{DeichmannGracien2018}. \REVISION{Indeed, while a sufficiently long and intense initializing laser pulse can in principle bring the NV spin population to a steady state without requiring a full thermal reset, the pulsed schemes remain limited by the moderate photon count rate per readout, which necessitates repeating the sequence up to a million times over a time window comparable to $T_1$ to achieve adequate signal-to-noise ratio}.
Together, pulsed $T_1$ measurements are most effective only within a narrow range of intermediate $T_1$ values, where the relaxation is neither too fast to resolve nor too slow to measure efficiently.

\REVISION{As an alternative to time-domain pulsed relaxometry (TDR), the spin relaxation can also be probed via continuous-wave optically detected magnetic resonance (cw-ODMR), whose signal, despite being more commonly used to sense static magnetic fields, also encodes information about spin dynamics.} \cite{saijo2018ac}. In cw-ODMR, the NV spins are simultaneously driven by continuous laser and microwave fields, enabling rapid measurements with high sensitivity and efficient noise rejection through lock-in detection \cite{clevenson2018robust}. Importantly, cw-ODMR naturally avoids the need for fast pulse switching and precise timing control, making it particularly attractive for nanoscale and ensemble-based measurements. In addition, cw-ODMR readily supports modulation of the microwave frequency or amplitude, allowing the fluorescence response to be demodulated at the modulation frequency. \REVISION{Crucially, when the modulation frequency approaches the intrinsic relaxation rate of a driven system, the population can no longer adiabatically follow the drive, causing the response to acquire a characteristic frequency dependence. This same principle may allow the fluorescence response of NV centers to reveal information about their spin relaxation dynamics.} Although modulation-based cw techniques are widely used to enhance sensitivity in nanoscale magnetometry and thermometry \cite{el2017optimised, ma2018magnetometry, singam2020nitrogen} and to measure the excited state lifetimes of metastable qubit states \cite{de2003recombination}, they have not yet been established as a quantitative and reliable method to directly extract the spin–lattice relaxation time $T_1$.

Here we address this gap by introducing a frequency-domain continuous-wave relaxometry (FDR) for extracting $T_1$ directly from cw-ODMR using low-frequency microwave amplitude modulation. By modulating the microwave field at frequency $\omega_{\mathrm{m}}$ and detecting the fluorescence with a lock-in amplifier, we obtain the complex frequency response $H(\omega_{\mathrm{m}})$. In a broad range of experimentally relevant regimes, this response is dominated by a single slow mode that can be accurately described by a first-order low-pass filter with an effective time constant $T_{\mathrm{eff}}$. We establish a direct connection between $T_{\mathrm{eff}}$ and the intrinsic spin–lattice relaxation time $T_1$, and demonstrate how the microwave-power dependence of $T_{\mathrm{eff}}$ enables $T_1$ to be quantitatively recovered from purely continuous-wave measurements. Experimentally, we validate the method by measuring $T_1$ in both bulk diamond and nanodiamond samples, obtaining excellent agreement with conventional pulsed protocols. Finally, we apply the technique to quantum relaxometry of biologically relevant paramagnetic ions in aqueous environments. Together, these results establish cw-ODMR modulation as a reliable, efficient, and broadly applicable route to measure $T_1$ across NV platforms ranging from bulk diamond to nanoscale ensembles, yielding fast and quantitative relaxation measurements in complex environments.

\section*{Theoretical background of FDR}

To enable efficient broadband measurements of $T_1$ time, we develop a frequency-domain relaxometry scheme based on the fundamental equivalence between time-domain relaxation dynamics and their frequency-domain response for the first-order kinetics \cite{depinna1984frequency}. Our approach is rooted in frequency-resolved spectroscopy, a well-established technique in which the in-phase ($X$) and quadrature ($Y$) components of the photocurrent detected by a lock-in amplifier are analyzed to extract photoluminescence lifetimes (see, for example, \cite{bort1991geminate,kaplan2010lifetime}). The FDR extends this framework to the dynamics of a ground-state spin qubit. We show that the microwave-dependent fluorescence intensity (ODMR signal) of NV centers exhibits a single effective decay constant, which approaches the ground-state spin relaxation time $T_1$ in the limit of low microwave and optical excitation powers.

\begin{figure*}[htp]
    \centering
    \begin{minipage}{0.42\textwidth}
        \includegraphics[width=1\linewidth]{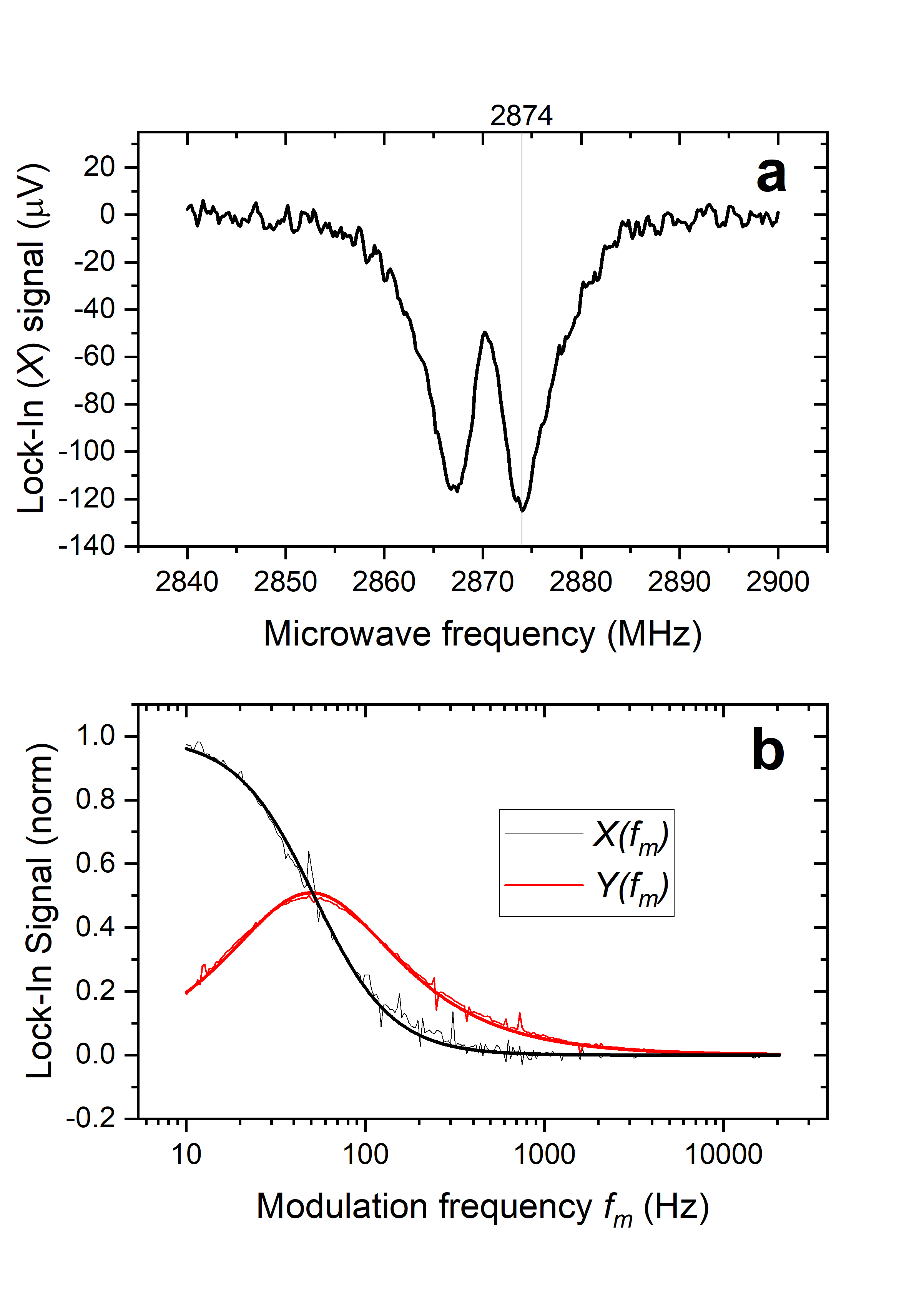}
    \end{minipage}
    \begin{minipage}{0.57\textwidth}
        \includegraphics[width=1\linewidth]{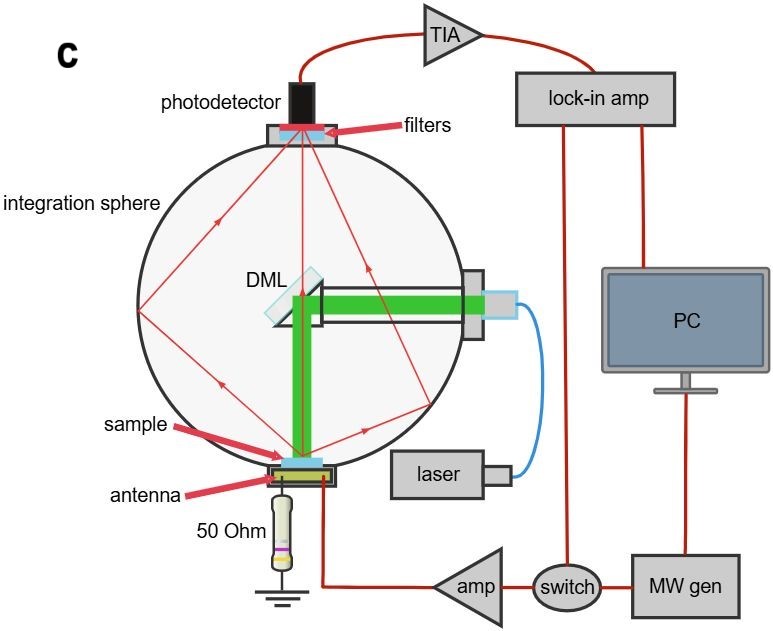}
    \end{minipage}
    \caption{A typical measurement protocol for the IN3x3 sample measured at \REVISION{-10~dBm} MW power and a laser intensity of 0.3~W/cm$^2$: (a) the cw-ODMR spectrum; (b) the $X$ and $Y$ components of the lock-in signal recorded at 2874~MHz as a function of the modulation frequency $f_{\mathrm{m}}$. \REVISION{The solid lines correspond to the fitting using equations~(\ref{eq5}) with $T_{\mathrm{eff}} = 3.2$~ms.} (c) Scheme of cw-ODMR setup.}
    \label{fig:1_FRS_IN}
\end{figure*}

To this end, we consider the relevant NV$^-$ transition (e.g. $|m_s=0\rangle \leftrightarrow |m_s=-1\rangle$) as an effective two-level system driven resonantly by a microwave field with Rabi frequency $\Omega$, in the presence of longitudinal relaxation characterized by $T_1$. The dynamics are governed by the Bloch equations, and the population $n$ of the state $|m_s=0\rangle$ evolves as follows \cite{CarringtonMcLachlan1967Ch1}:

\begin{equation}
    \frac{dn}{dt}=-2Pn-\frac{n-n_0}{T_1}, \label{eq1}
\end{equation}
where $n_0$ denotes the population immediately after optical polarization and $P$ is the microwave-induced transition probability, proportional to the square of the Rabi frequency ($P\propto \Omega^2$). Solving equation~(\ref{eq1}), and noting that the ODMR signal corresponds to a fluorescence change $S(t)$ proportional to the population difference, we obtain after discarding stationary contributions 
\begin{equation}
    \Delta S(t) \approx \frac{2PT_1}{1+2PT_1} e^{-(\frac{1}{T_1}+2P)t}. \label{eq2}
\end{equation}
Under continuous wave conditions and small microwave amplitude modulation $\delta \Omega$ at frequency $\omega_{\mathrm{m}}$, the fluorescence $S(t)$ can be linearized:
\begin{equation}
  \delta S(t) \approx \Re[\delta \Omega H(\omega_{\mathrm{m}}) e^{i \omega_{\mathrm{m}} t}],  \label{eq3}
\end{equation}
where $H(\omega_{\mathrm{m}})$ is the complex frequency response of the spin system. Although $H(\omega_{\mathrm{m}})$ in general contains several relaxation modes \REVISION{including microwave and laser contributions, it is fundamentally limited by spin–lattice relaxation.}
%, in many regimes the response is dominated by a single slow mode associated with spin–lattice relaxation. 
In this limit, the system is well described by a first-order response function and the lock-in components are
\begin{equation}
X(\omega_{\mathrm{m}}) \propto  \frac{1}{1 + (\omega_{\mathrm{m}} T_{\mathrm{eff}})^2}, \hspace{5mm}
Y(\omega_{\mathrm{m}}) \propto  -\frac{\omega_{\mathrm{m}} T_{\mathrm{eff}}}{1 + (\omega_{\mathrm{m}} T_{\mathrm{eff}})^2}, \label{eq5}
\end{equation}
where $T_{\mathrm{eff}}=T_1/(1+2PT_1)$ is an effective time constant. This is the characteristic response of a first-order low-pass filter with cutoff frequency $\omega_c = 1/T_{\mathrm{eff}}$.  
The effective time constant $T_{\mathrm{eff}}$ can be extracted directly from the frequency response. Experimentally, it is obtained either from the half-decay point of the in-phase component $X(\omega_{\mathrm{m}})$ or, more conveniently, from the frequency $f_{\mathrm{m}}$ at which the quadrature component $Y(\omega_{\mathrm{m}})$ reaches its maximum $T_{\mathrm{eff}}=\frac{1}{2\pi f_m}$.

Beyond the intrinsic longitudinal relaxation time $T_1$, the effective time $T_{\mathrm{eff}}$ incorporates additional drive-dependent rates. Using the definition from equation~(\ref{eq5}) and considering continuous-wave optical pumping, a rate-equation treatment yields, to leading order,
\begin{equation}
    \frac{1}{T_{\mathrm{eff}}} \approx \frac{1}{T_1} + 2P + \Gamma_p. \label{eq_21}
\end{equation}
Here $\Gamma_p = \sigma_{\lambda}I/(h\nu)$ \cite{jeske2017stimulated} is the polarization rate of stimulated emission, where $I$ is the power per area of the light field, ($h\nu$) is the corresponding photon energy, and $\sigma_{\lambda}$ is the stimulated emission cross-section at a given wavelength $\lambda$. Since $P \propto \Omega^2 \propto P_{\mathrm{MW}}$, the microwave power and $\Gamma_p \propto I$, the equation (\ref{eq_21}) can be rewritten as
\begin{equation}
    \frac{1}{T_{\mathrm{eff}}(P_{\mathrm{MW}}, P_L)}
\approx \frac{1}{T_1} + \alpha P_{\mathrm{MW}} + \beta I. \label{eq_22}
\end{equation}
By measuring $T_{\mathrm{eff}}$ as a function of $P_{\mathrm{MW}}$ and $I$ and extrapolating to the vanishing microwave and laser power, we can extract intrinsic "dark" relaxation time $T_1$ in the frequency domain without explicit time-resolved measurements. 

%\section{Results}
\section*{Validation of FDR in bulk samples}

Firstly, to validate the proposed FDR method and benchmark it against time-domain relaxometry, we investigated two single-crystal diamond samples grown by chemical vapor deposition (CVD), containing different concentrations of NV$^-$ centres. The measurement protocol is summarized in Fig.~\ref{fig:1_FRS_IN}(a,b). We begin by recording the cw-ODMR spectrum and determining the precise positions of the spin resonances (see Fig.~\ref{fig:1_FRS_IN}(a) and Supplementary Fig.~1 for the cw-ODMR spectrum of the DNV sample). In both samples, the ODMR response exhibits the characteristic double-dip (double-well) structure typical of NV$^-$ ensembles at close to 2.87~GHz, reflecting the two allowed spin transitions within the ground-state triplet manifold. 

Once the resonance frequencies are accurately identified, the microwave excitation is fixed at the frequency corresponding to the maximum lock-in contrast. At this operating point, we record the $X$ and $Y$ components of the lock-in amplifier while sweeping the modulation frequency $f_m$ over a broad range, as shown in Fig.~\ref{fig:1_FRS_IN}(b). Starting from a near-plateau around 10~Hz, the $X(f_m)$ component exhibits a gradual roll-off, enabling a reliable extraction of the characteristic frequency from the half-power point. In turn, the analysis of the $Y(f_m)$ component is even more straightforward, as it reaches a maximum at the characteristic frequency, corresponding to a 45 degree phase lag of the spin response; thus the effective relaxation time can be determined from equation~(7). Both $X$ and $Y$ components can be fitted with the corresponding analytical expressions \cite{depinna1984frequency}, allowing us not only to extract the characteristic time constants but also to quantify the associated uncertainties. Notably, the experimental $X(f_{\mathrm{m}})$ and $Y(f_{\mathrm{m}})$ curves in Fig.~\ref{fig:1_FRS_IN}(b) identify the same characteristic modulation frequency, providing a direct and model-independent consistency check of the method.

\begin{figure*}[ht]
    \centering
    \begin{minipage}{0.32\textwidth}
        \includegraphics[trim=10mm 0mm 20mm 0mm, clip, width=\linewidth]{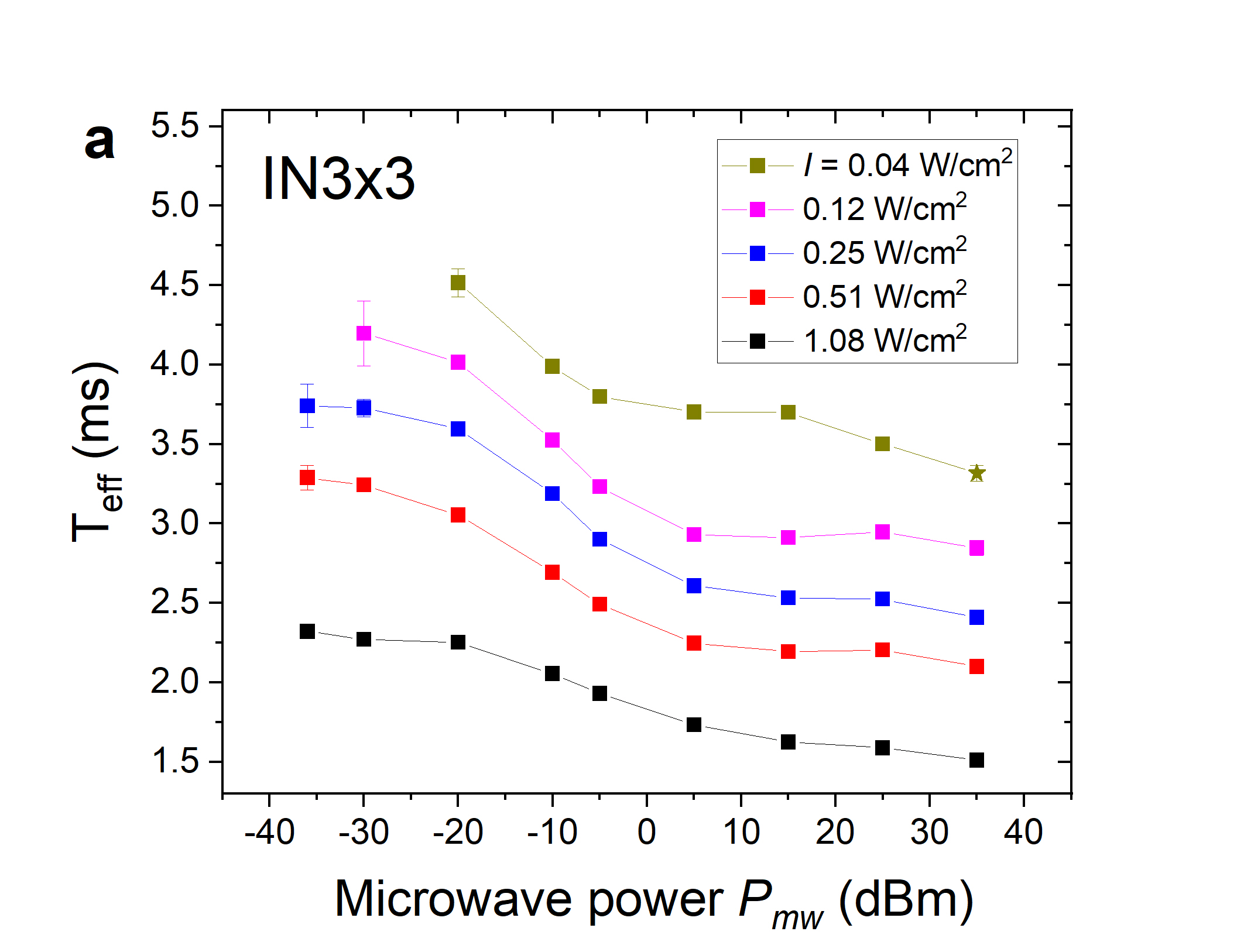}
    \end{minipage}
    \hfill
    \begin{minipage}{0.32\textwidth}
        \includegraphics[trim=10mm 0 20mm 0, clip, width=\linewidth]{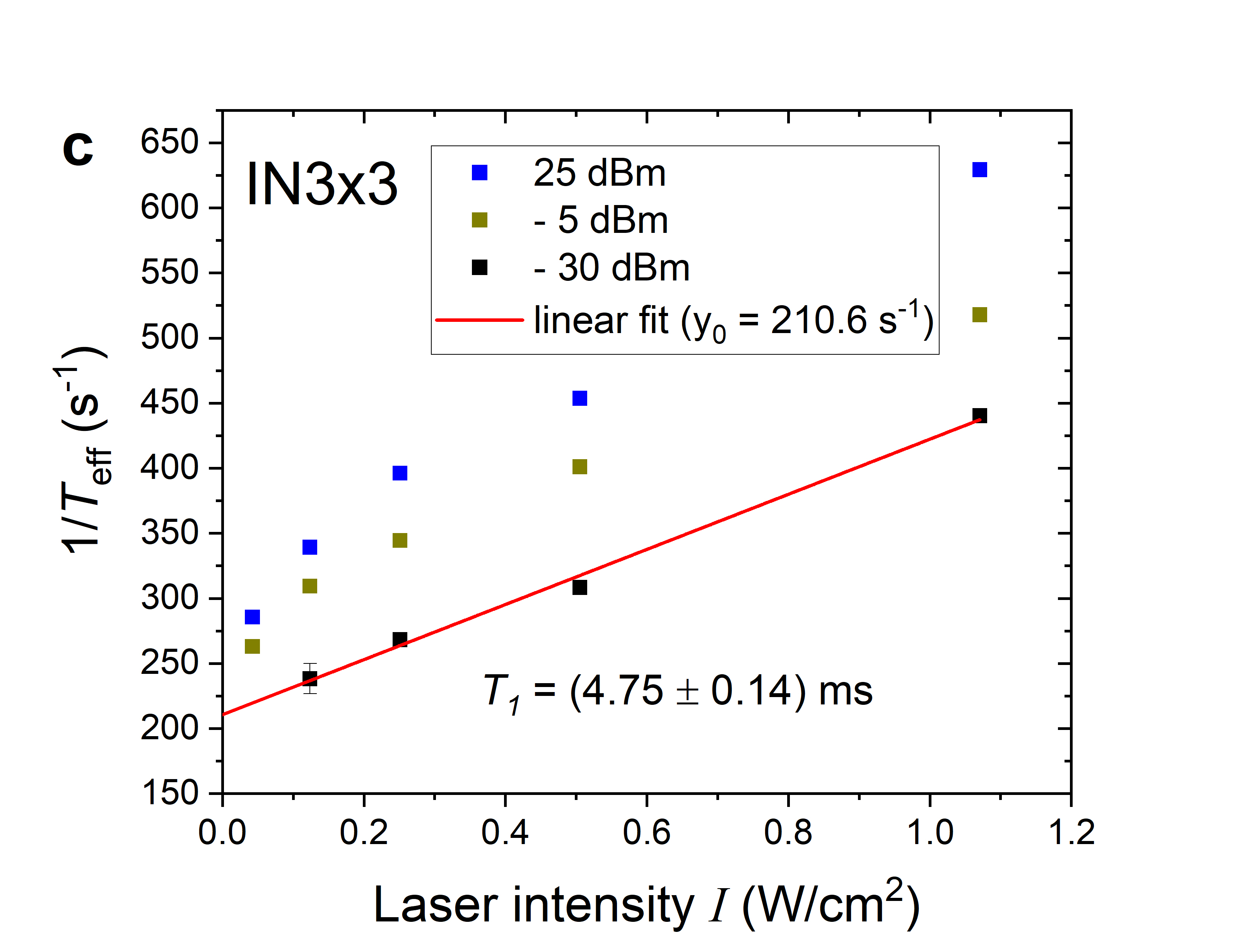}
    \end{minipage}
    \hfill
    \begin{minipage}{0.32\textwidth}
        \includegraphics[trim=10mm 0 20mm 0, clip, width=\linewidth]{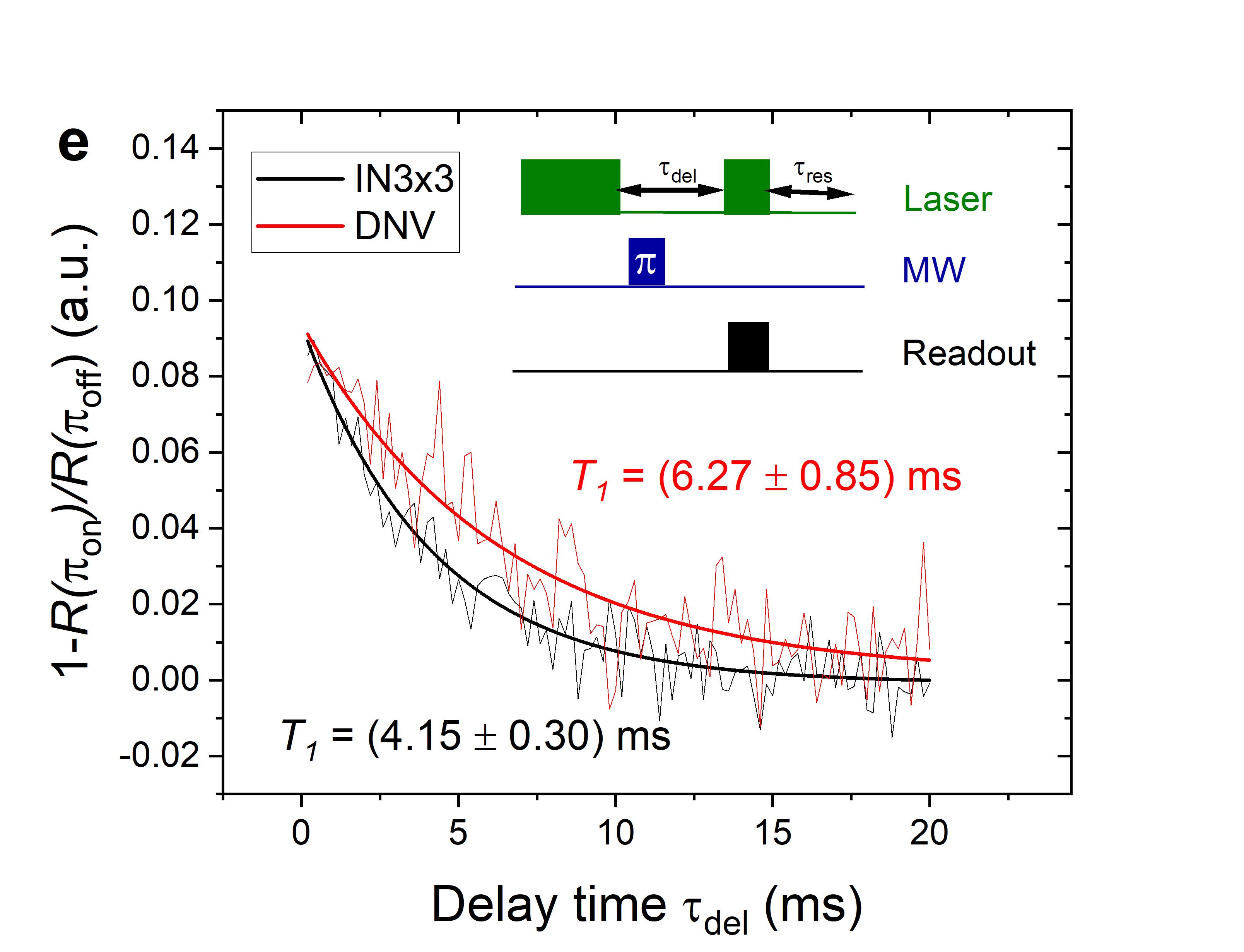}
    \end{minipage}
    \begin{minipage}{0.32\textwidth}
        \includegraphics[trim=10mm 0mm 20mm 0mm, clip,width=\linewidth]{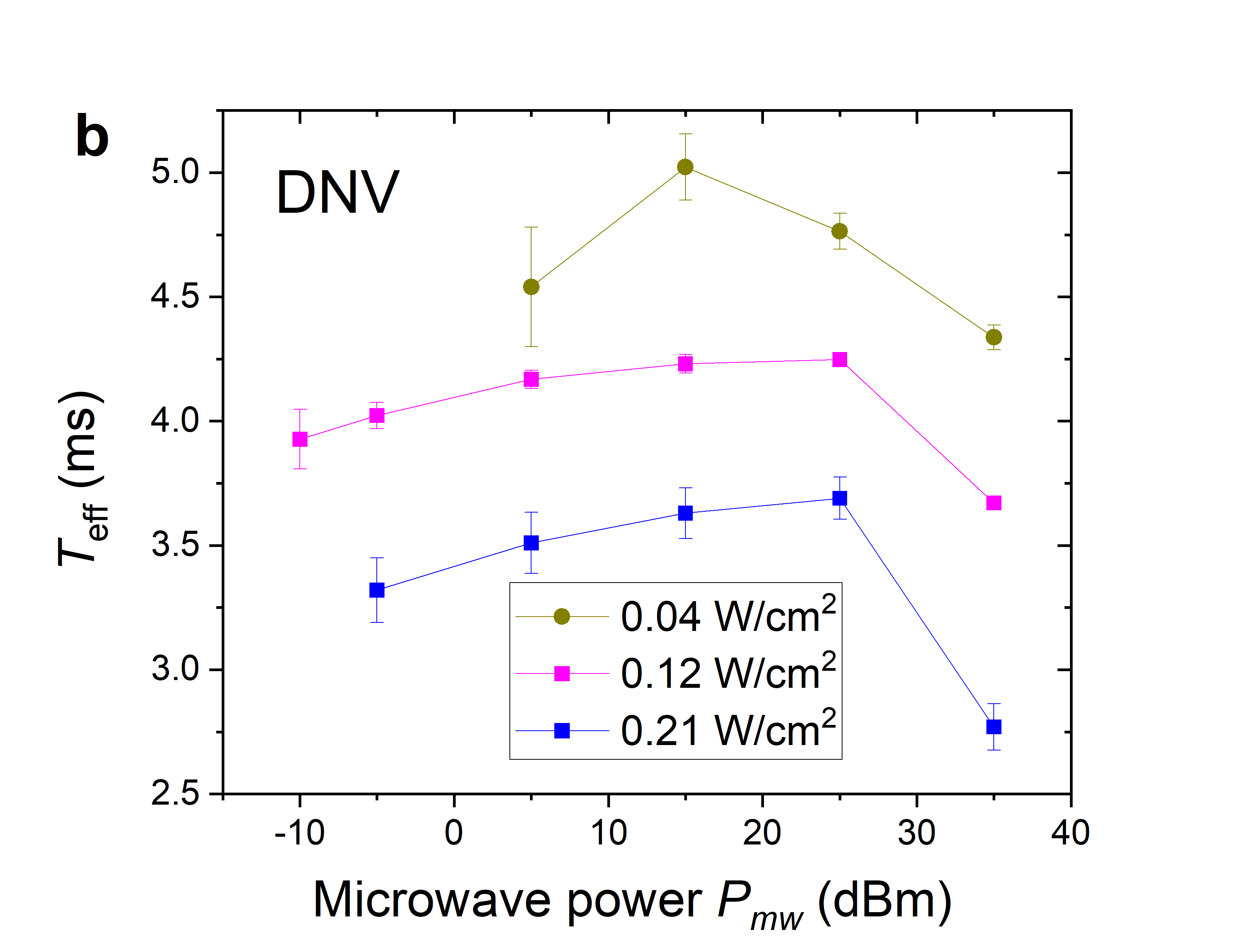}
    \end{minipage}
    \hfill
    \begin{minipage}{0.32\textwidth}
        \includegraphics[trim=10mm 0 20mm 0, clip,width=\linewidth]{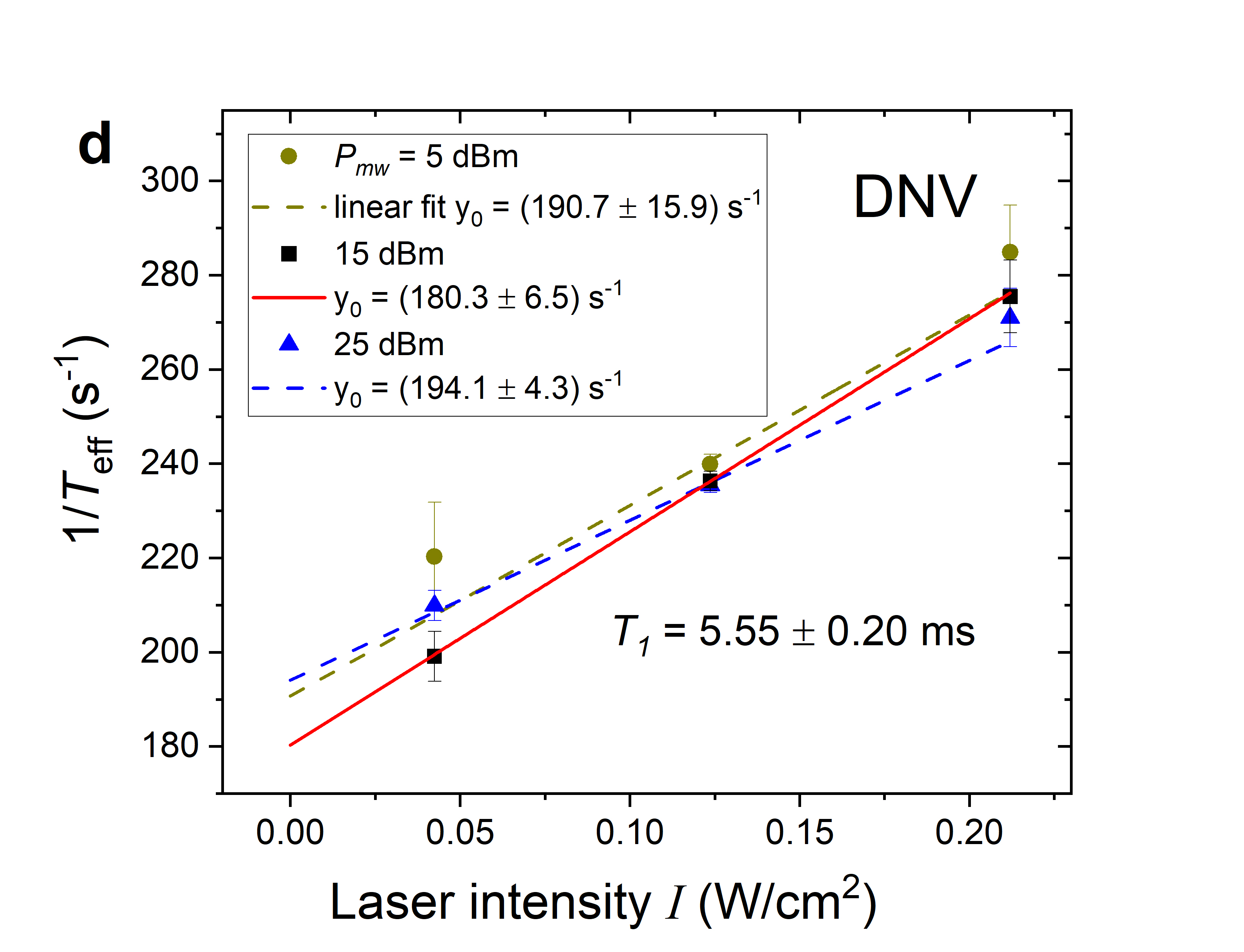}
    \end{minipage}
    \hfill
    \begin{minipage}{0.32\textwidth}
        \includegraphics[trim=10mm 0 20mm 0, clip,width=\linewidth]{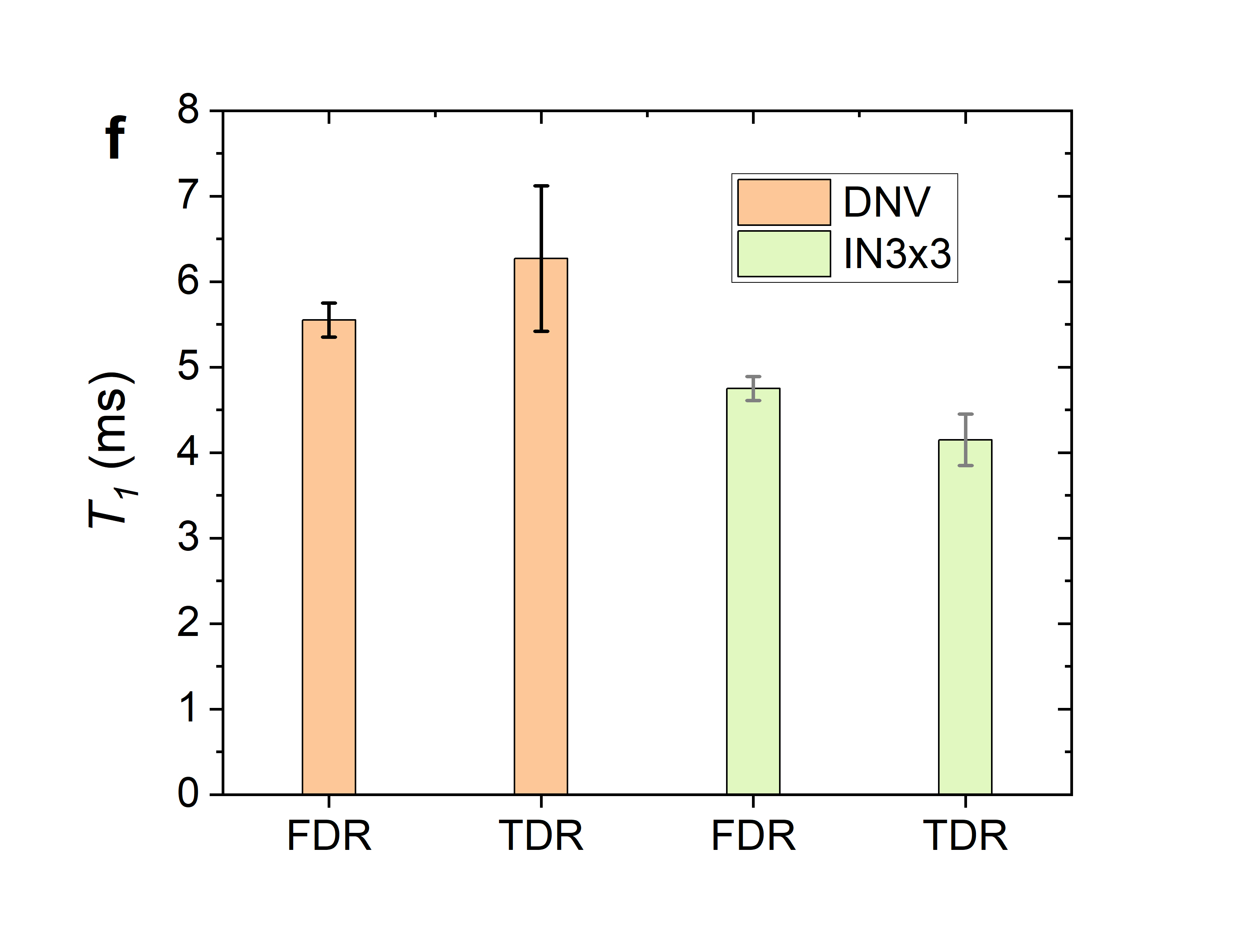}
    \end{minipage}
    \caption{Results of the $T_1$ measurements in bulk samples. Dependence of the effective time $T_{\mathrm{eff}}$ on the microwave radiation power at different laser intensities $I$ for IN3x3 (a) and DNV (b) samples. Dependence of the effective rate $1/T_{\mathrm{eff}}$ on the laser intensity at different microwave power $P_{mw}$ for IN3x3 (c) and DNV (d) samples. The time $T_1$ is calculated based on linear approximations to zero laser power, shown by solid red lines. (e) Results of $T_1$ measurement for IN3x3 (black) and DNV (red) samples using TDR protocol (inset). 
    %The NV centers are initially polarized into the $|m_s = 0\rangle$ state by an optical pulse lasting 1~ms. Next, a microwave $\pi$-pulse are applied, typically for 250~ns. Next, after the free evolution time $\tau_{\mathrm{del}}$, a second optical pulse is applied for the readout ($R$) lasting 5~$\mu$s. The system is then allowed to fully relax over time $\tau_{\mathrm{res}}$. After this, the same sequence is applied, but with the $\pi$-pulse omitted, and the resulting time  $T_1$ is derived from the exponential decay of the $1-R(\pi_\text{on})/R(\pi_\text{off})$ function. 
    (f) Comparison between FDR and TDR measurement methods for IN3x3 and DNV samples. All errors here and throughout were determined based on fitting using analytical expressions.}
    \label{fig:bulk}
\end{figure*}

According to equation~(\ref{eq_22}), the measured effective relaxation time $T_{\mathrm{eff}}$ includes contributions from both the applied microwave field and laser-induced spin pumping, whereas our goal is to determine the intrinsic, “dark” relaxation time $T_1$. Using the protocol described above, we studied the dependence of $T_{\mathrm{eff}}$ on microwave and laser powers for both samples (Fig.~\ref{fig:bulk}(a–d)). For the IN3x3 sample (Fig.~\ref{fig:bulk}(a)), $T_{\mathrm{eff}}$ evolves nonlinearly with microwave power at all laser intensities, \REVISION{reaching a plateau} at a relatively low power ($-30$~dBm). In contrast, for the DNV sample (Fig.~\ref{fig:bulk}(b)), \REVISION{the plateau} occurs at a higher microwave power ($P_\mathrm{mw}=25$~dBm), \REVISION{due to its larger size, which massively reduces the effective microwave power experienced by the spins.} This nonlinear dependence on microwave power at high excitation intensities follows directly from equation~(\ref{eq2}). The internal dynamics of cw-ODMR becomes nonlinear as the spin transition approaches saturation, so that the steady-state fluorescence contrast and relaxation rates are no longer proportional to the microwave power. In this regime, the simple low-power approximation (equation~\ref{eq_22}) is no longer valid. Furthermore, in Figs.~\ref{fig:bulk}(c,d) the effective relaxation rates $1/T_{\mathrm{eff}}$ are plotted as a function of laser intensity $I$ for different microwave powers. As seen in Fig.~\ref{fig:bulk}(c)\REVISION{, for the IN3x3 sample}, when system is in the MW saturation regime, the laser power dependence of $1/T_{\mathrm{eff}}$ is sub-linear. In turn, achieving a linear dependence of $1/T_{\mathrm{eff}}$ on $I$ requires operating at lower microwave powers where $T_{\mathrm{eff}}$ approaches the plateau. \REVISION{Meanwhile, for the DNV sample (Fig.~\ref{fig:bulk}(d)) the dependence of $1/T_{\mathrm{eff}}$ on the laser power is linear over a wide range of microwave radiation powers below 25~dBm.} Extrapolating this linear trend to zero laser intensity then yields the intrinsic relaxation times: $T_1=(4.75\pm 0.14)$~ms for IN3x3 and $T_1=(5.55\pm 0.20)$~ms for DNV (Fig.~\ref{fig:bulk}(c,d)).

To confirm these results, we measured $T_1$ using a time-domain pulse protocol. Notably, pulsed $T_1$ measurements are highly sensitive to the details of the pulse scheme, with the longest relaxation times obtained using the optimized sequence shown in Fig.~2(e) and described in details in the Methods section. The corresponding measurement results are also presented in the same figure, revealing $T_1=(4.15 \pm 0.30)$~ms for the IN3x3 sample and $T_1=(6.27 \pm 0.85)$~ms for the DNV sample. Fig.~\ref{fig:bulk}(f) summarizes the cumulative $T_1$ results obtained with both methods. For the DNV sample, the values are fully consistent within the experimental error. For the IN3x3 sample, the pulsed $T_1$ is slightly shorter than the FDR result, which likely reflects suboptimal pulse settings. In particular, accurately determining the $\pi$-pulse length in this sample is complicated by the shortened transverse relaxation time $T_2^*$, caused by strong spin-spin interactions at high NV$^-$ concentration, see Supplementary Note~1.

\begin{figure}
    \centering
    \includegraphics[width=0.5\linewidth]{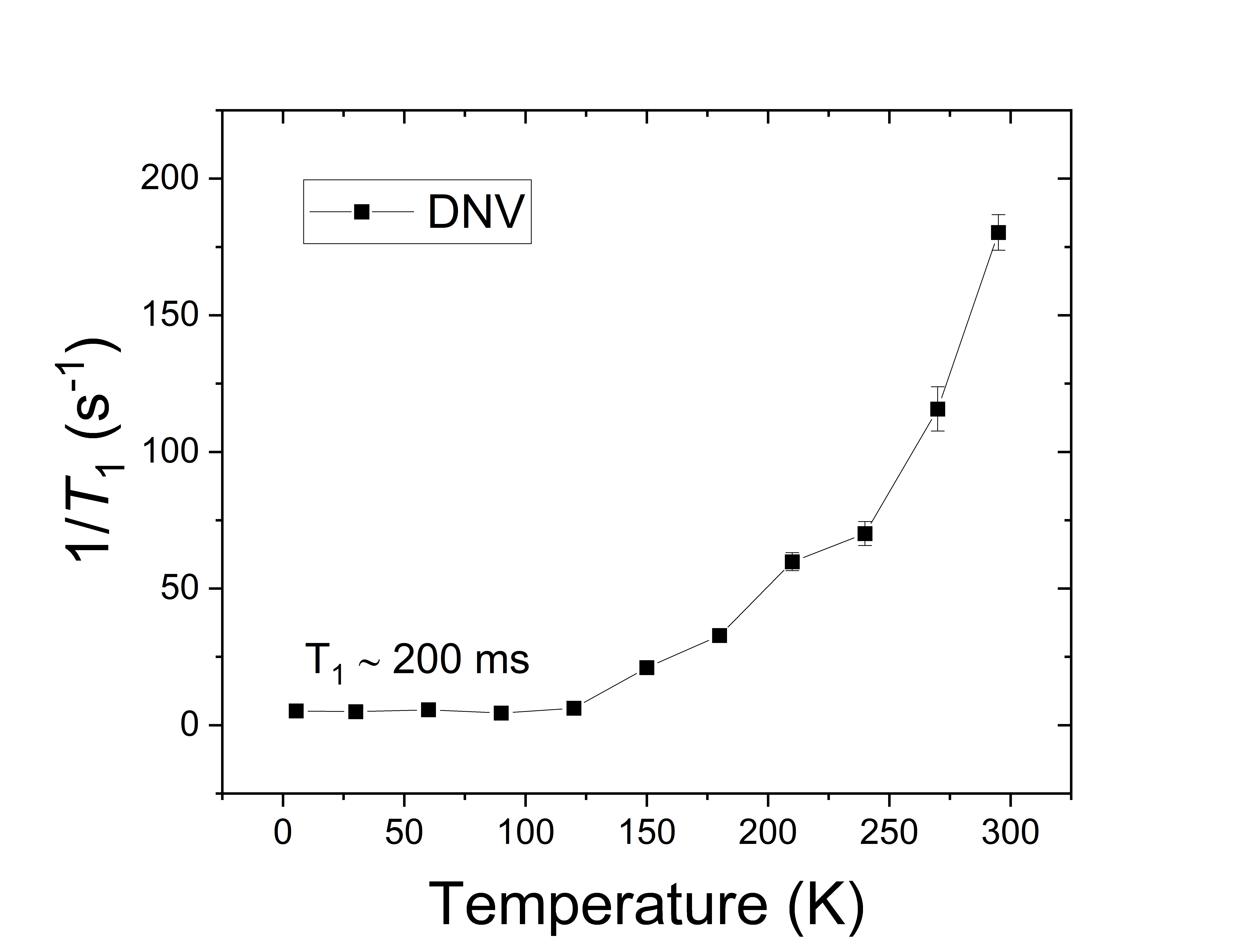}
    \caption{Temperature dependence of the relaxation rate measured by the FDR method in the DNV sample.}
    \label{fig:low_DNV}
\end{figure}

Interestingly, the measurement error for the pulsed TDR method is two to four times larger than for the FDR method, despite the total experiment time being at least six times longer in this specific case. This effect becomes especially critical when probing very long relaxation times, as each pulse sequence requires waiting for full spin relaxation. By contrast, in the FDR method, extending the measurement to long relaxation times primarily shifts the relevant modulation frequency band to lower values, without a proportional increase in total measurement time. To illustrate this point, Fig.~\ref{fig:low_DNV} presents the temperature dependence of the relaxation rate $1/T_1$ in a DNV sample measured using the FDR method. As shown, the relaxation rate decreases sharply upon cooling and saturates below  $\sim$120~K. This behavior was unambiguously identified in prior work as arising from phonon contributions \cite{cambria2023temperature}, further reinforcing our claim that the measured relaxation time is indeed the spin-lattice relaxation $T_1$. Moreover, at cryogenic temperatures the $T_1$ time reaches values of $\sim$200 ms. Determining such long relaxation times becomes challenging for the TDR measurements with the present pulse sequence, see the results of this experiment at 10~K in Supplementary Fig.~3. Specifically, even after 48-hour measurements, the signal-to-noise ratio (SNR) is around 1, which does not allow obtaining reliable values of the relaxation time. Although the FDR approach is also susceptible to technical noise, such as low-frequency drift and $1/f$ noise in the sub-hertz range, these effects can be effectively mitigated by increasing the laser power during extrapolation, thereby maintaining sufficient SNR. We therefore expect that the FDR approach can be broadly applied beyond diamond samples, particularly to silicon carbide, where the $T_1$ times of divacancy and transition-metal defects can extend to the minute scale \cite{anderson2022five,PhysRevApplied.22.044078}.

\section*{Application of FDR to nanodiamond samples}

\begin{figure}
    \centering
    \includegraphics[width=0.5\linewidth]{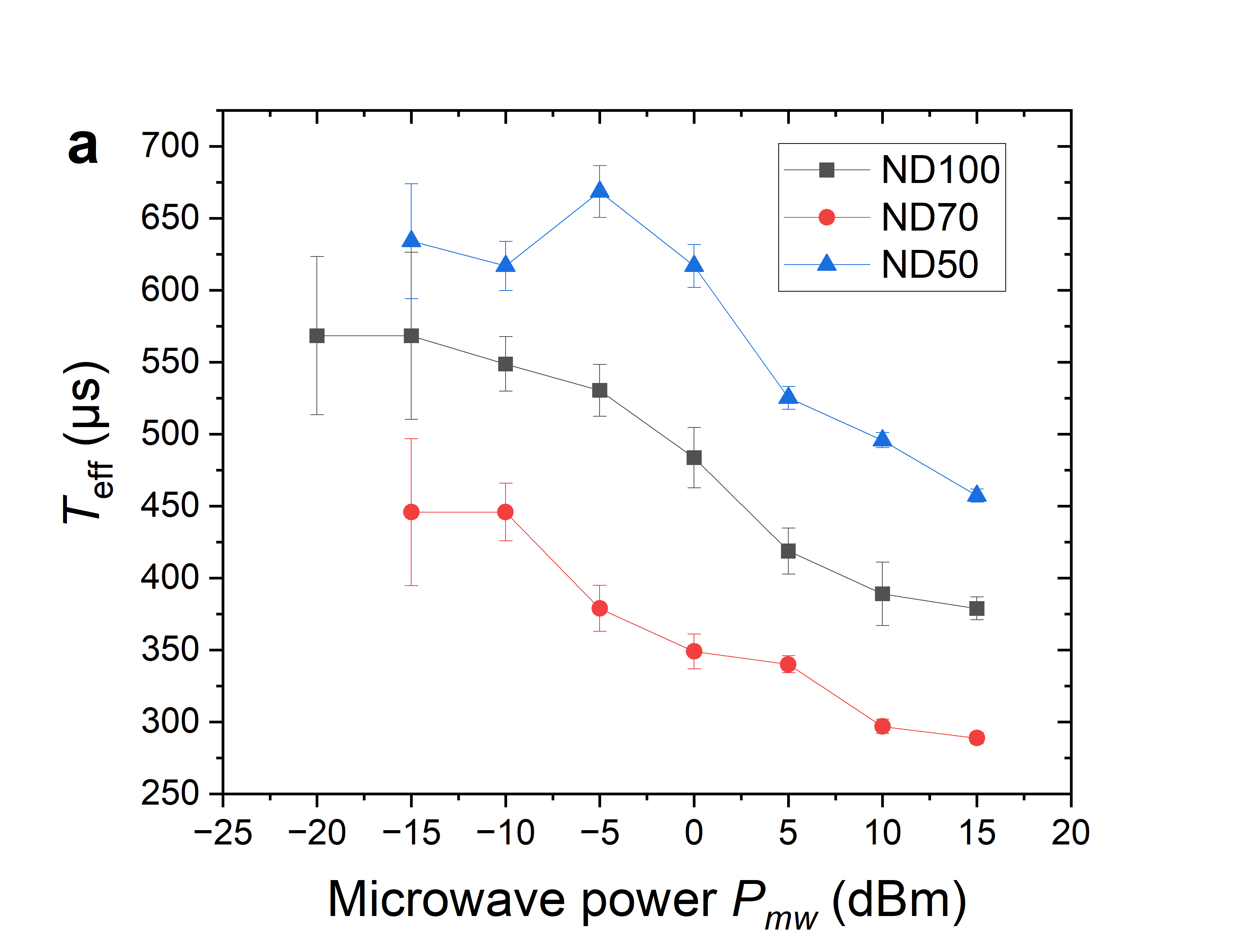}
    \includegraphics[width=0.5\linewidth]{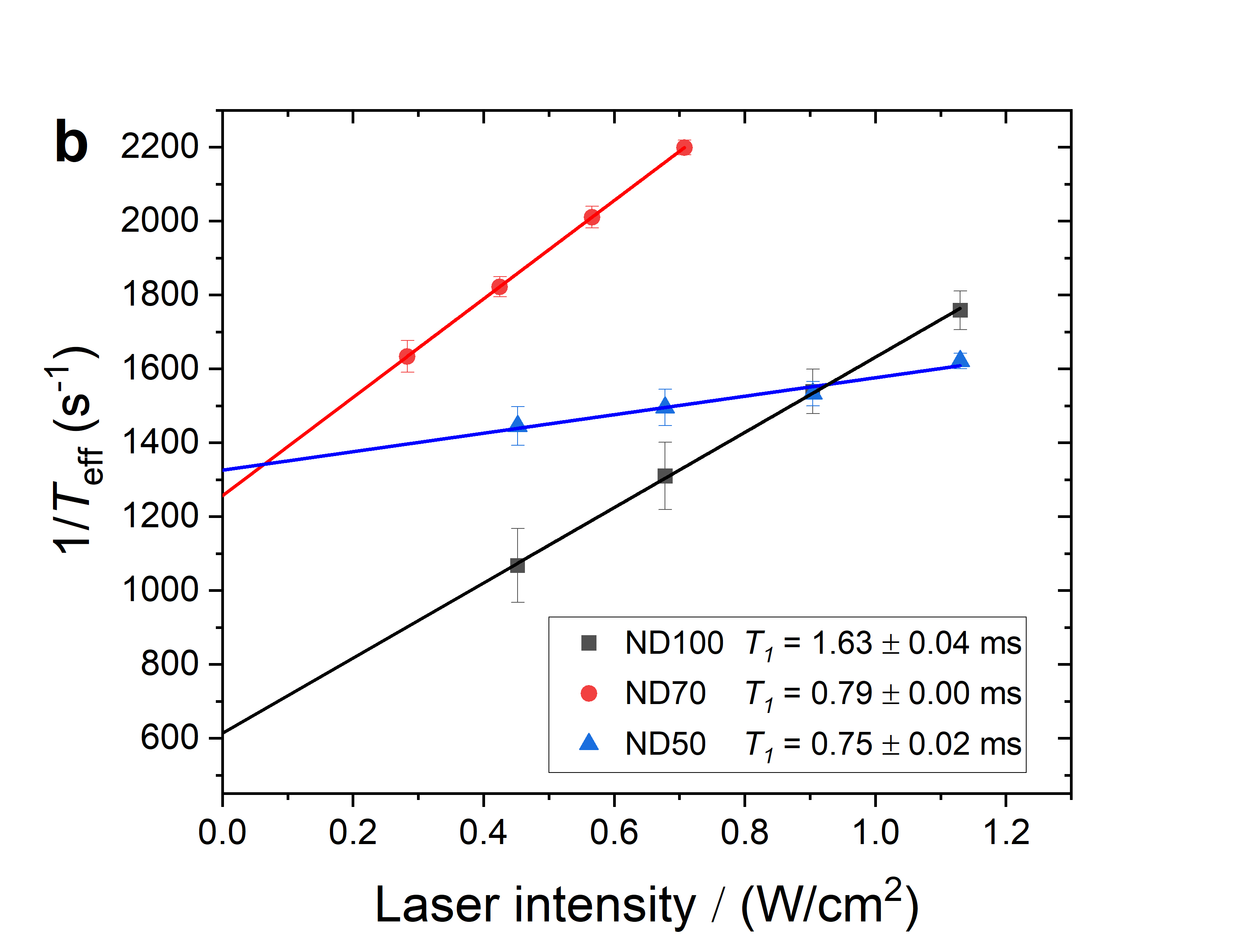}
    \caption{(a) Dependence of the effective time $T_{\mathrm{eff}}$ on the microwave radiation power for different sizes of nanodiamonds. (b) Dependence of the effective rate $1/T_{\mathrm{eff}}$ on the laser intensity for different sizes of nanodiamonds. The $T_1$ time is calculated based on linear approximations to zero laser power, shown by solid lines.}
    \label{fig:ND_MW_LP}
\end{figure}

Having demonstrated that our FDR method accurately measures $T_1$ relaxation times in bulk samples, we next investigate NV$^-$ centres in nanodiamonds of biologically relevant size (below 100 nm). Conventional pulsed schemes are less effective in these systems, as multi-exponential fitting of weak, noisy signals can introduce large errors \cite{grant2023method}, whereas the FDR approach is expected to perform best. The ODMR spectra of the ND100 are shown in Supplementary Fig.~5; spectra for the other samples are qualitatively similar and described in detail elsewhere \cite{jegenyes2025materials,mzyk2022relaxometry}.
Fig.~\ref{fig:ND_MW_LP}(a) shows the dependence of $T_1$ on the applied MW field for three representative nanodiamond sizes. In each case, the dependence exhibits a saturation regime similar to that observed in bulk samples, consistent with MW-induced relaxation dominating at high drive strengths. In the linear MW power regime, we further analyse the dependence of the effective relaxation rate ($1/T_{\mathrm{eff}}$) on laser power. As shown in Fig.~\ref{fig:ND_MW_LP}(b), $1/T_{\mathrm{eff}}$ increases linearly with laser power for all three sizes. Extrapolation yields intrinsic $T_1$ values of 1.6~ms, 0.8~ms and 0.7~ms for ND100, ND70 and ND50, respectively, which are among the longest reported for nanodiamonds. In particular, relaxation times exceeding 1~ms have previously been observed only in isotopically purified materials or after surface modification \cite{oshimi2024bright,barzegaramiriolya2025functionalized}, whereas typical measurements report $T_1$ values of only a few hundred microseconds. 
Notably, the reduced relaxation times in smaller nanodiamonds shift the onset of saturation to higher MW fields (Fig.~\ref{fig:ND_MW_LP}(a)), resulting in an unexpected ordering of $T_{\mathrm{eff}}$ with nanodiamond size at elevated drive strengths. The comparatively long relaxation times measured here suggest that the intrinsic $T_1$ of nanodiamond NV centres may lie much closer to bulk values than commonly assumed, with shorter values in earlier reports likely reflecting limitations of non-optimised pulsed measurement protocols.
We note that the ensemble of nanodiamond NV spins resides in a substantially inhomogeneous environment, which makes it difficult, if not impossible, to achieve an optimised pulse scheme.

\begin{figure*}
    \centering
    \begin{minipage}{0.49\textwidth}
        \includegraphics[width=1\linewidth]{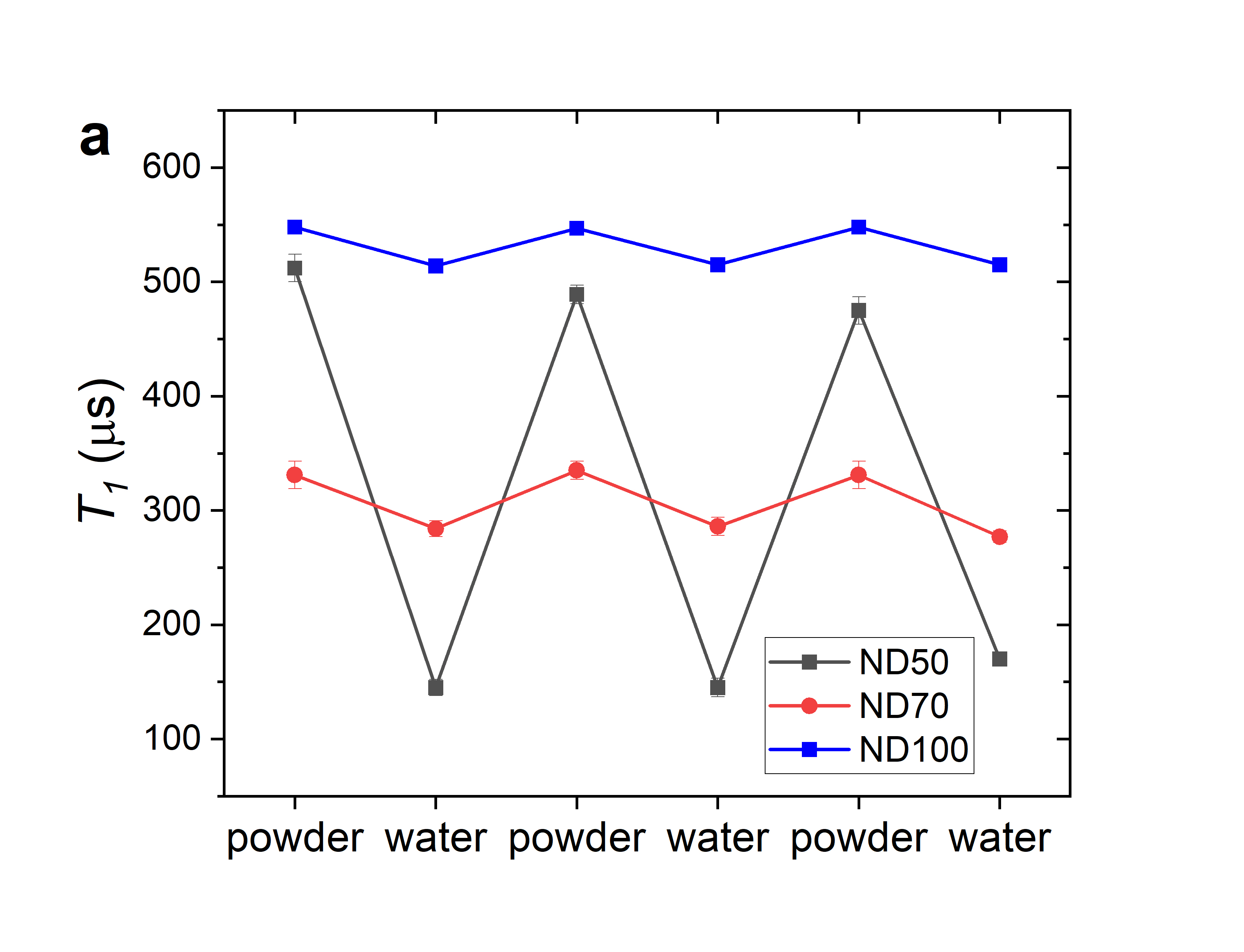}
    \end{minipage}
    \hfill
    \begin{minipage}{0.49\textwidth}
        \includegraphics[width=1\linewidth]{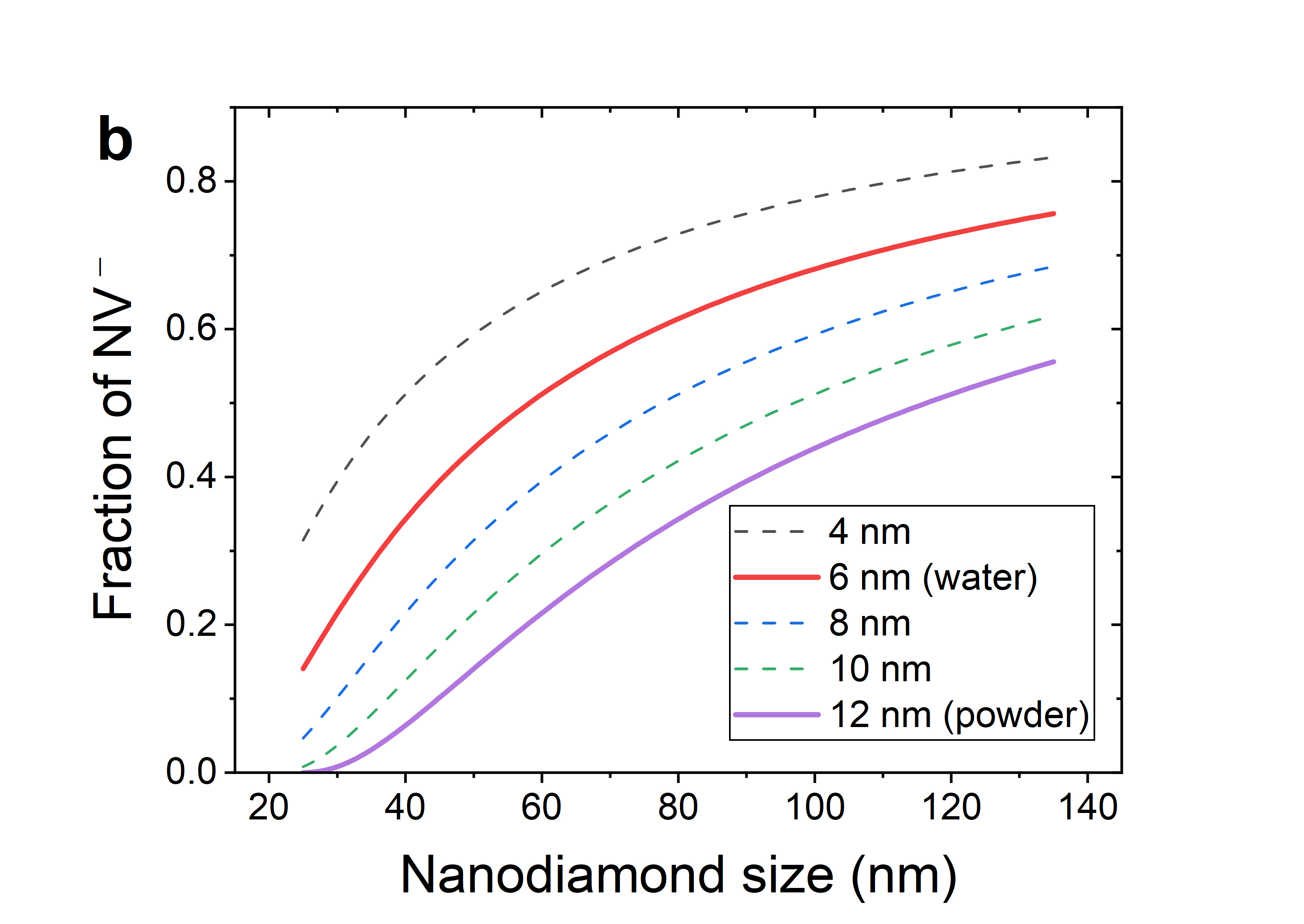}
    \end{minipage}
    \vfill
    \begin{minipage}{0.49\textwidth}
        \includegraphics[width=1\linewidth]{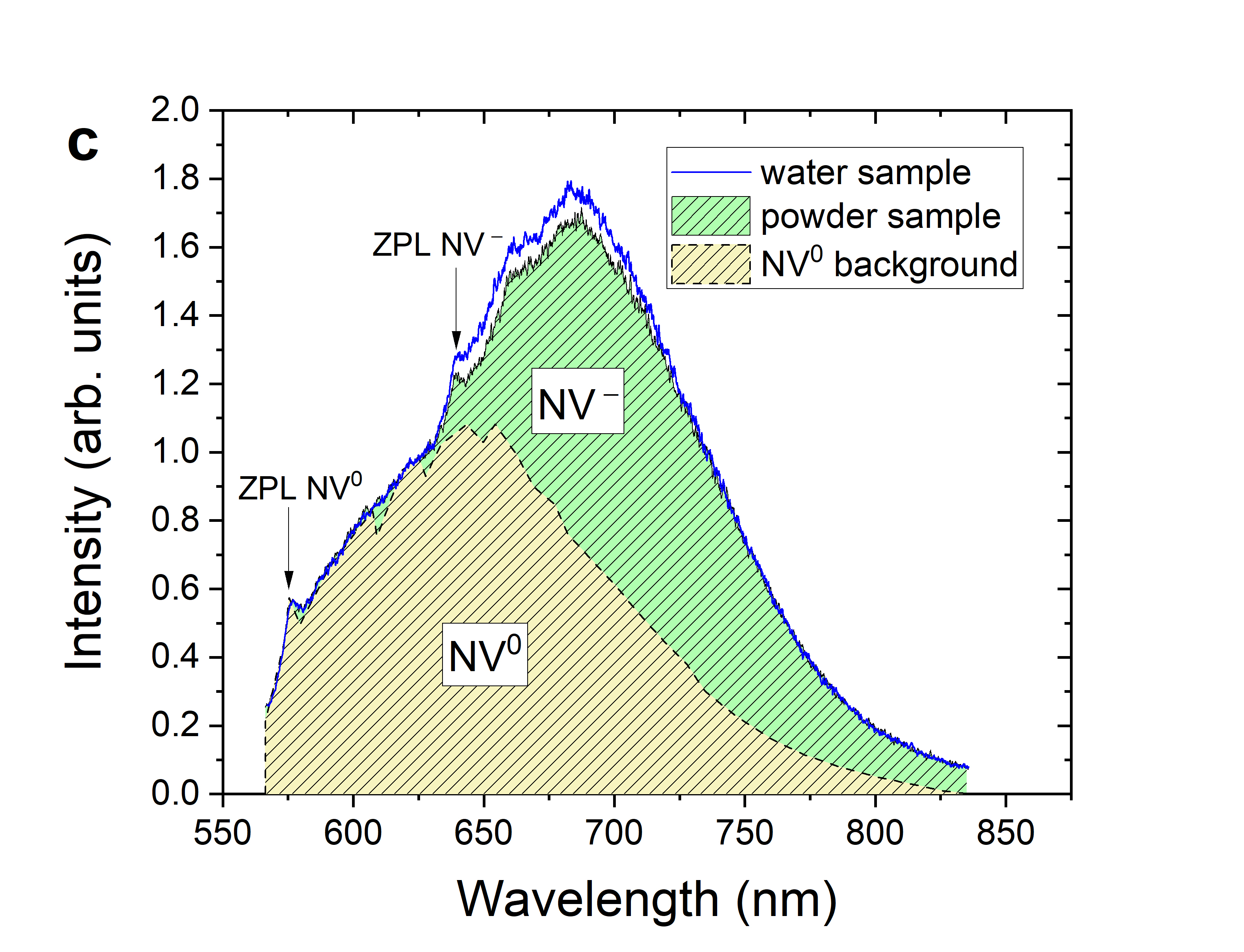}
    \end{minipage}
    \hfill
    \begin{minipage}{0.49\textwidth}
        \includegraphics[width=1\linewidth]{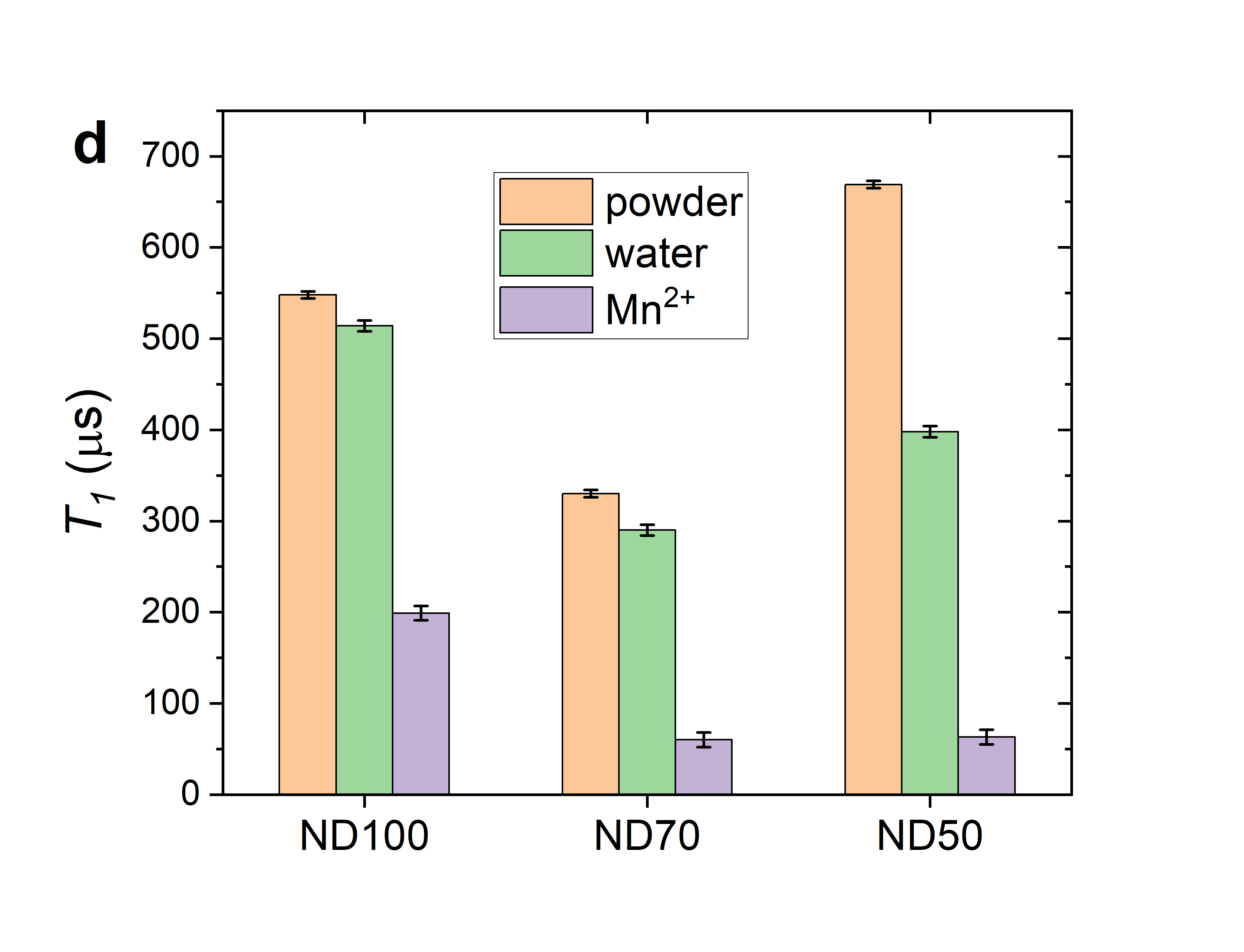}
    \end{minipage}
    \caption{(a) Results of comparative measurements of $T_1$ time for aqueous and dry samples of three types of nanodiamonds. (b) Simulated fraction of stable NV$^-$ centers in spherical nanodiamonds as a function of particle size. Solid curves show the fraction of NV$^-$ in water and powder (UHV), calculated using a purely geometrical model that accounts for near-surface instability as the ratio of stable-to-total volume. The indicated instability depths were obtained based on density functional theory calculations \cite{neethirajan2023controlled}. Dashed curves represent intermediate cases corresponding to variable experimental conditions.  (c) Photoluminescence spectra for ND50 (in water and powder) aligned with the NV$^0$ background from \cite{aslam2013photo} (d) Results of optimised FDR measurements in nanodiamonds in (citrus punch) powders,  (lime green) water samples and (pale purple) in the presence of Mn$^{2+}$ ions. \REVISION{Note that the same set of nanoparticles was used in every series.}}
    \label{fig:ND_H20}
\end{figure*}

One of the key applications of quantum relaxometry with fluorescent nanodiamonds is bio-sensing of spin markers in aqueous media \cite{grant2023method}, where we further tested our FDR method. Specifically, we aim to observe the response to dissolved Mn$^{2+}$ ions ($S$=5/2) \cite{ziem2013highly} with the goal of demonstrating sensitivity to micromolar concentrations. Importantly, in this approach, the focus is on measuring the difference in $T_1$ rather than the absolute values. For the nanodiamonds shown in Fig.~\ref{fig:ND_MW_LP}, the errors in the $T_1$ values significantly increase at low MW and laser powers due to the reduced signal strength compared to bulk samples. Consequently, we opted to perform "one-shot" measurements at moderate MW and laser powers, %which provide improved SNR, assuming that $\Delta T_{\mathrm{eff}} \approx \Delta T_1$ remains valid in this regime.
where $T_{\mathrm{eff}}$ is dominated by $T_1$ and SNR remains high. Prior to investigating the impact of manganese ions, we quantified the response of $T_1$ to water (Fig.~\ref{fig:ND_H20}(a)). By repeatedly adding and evaporating water, we observe  a systematic reduction of $T_1$ in the presence of water in all samples, with periodic and reproducible changes over successive cycles. The magnitude of the decrease is size dependent, ranging from $\sim6\pm2$\% for ND100 to $\sim10\pm4$\% for ND70 and $\sim50\pm10$\% for ND50 across different field strengths and samples. We attribute this behavior to a water-induced revival effect  \cite{neethirajan2023controlled}, stabilizing shallow NV$^-$ centers with reduced $T_1$. The simulation results in Fig.~\ref{fig:ND_H20} confirm that smaller ND50 particles are highly susceptible to this effect.  %In earlier work, we reported that the conversion of NV$^-$ to NV$^0$ occurs at depths of $\sim$12~nm in powder (under ultra-high vacuum conditions), whereas in aqueous environments this threshold shifts to $\sim$6~nm \cite{neethirajan2023controlled}. For a spherical nanodiamond of diameter $D$, the density of potential defect sites increases cubically toward the surface, setting a characteristic mean depth of NV centers on the order of $D/8$. Consequently, the smaller ND50 particles are inherently more susceptible to surface-mediated effects, as reflected in Fig.~\ref{fig:ND_H20}(b). Specifically, in this case the NV$^-$ fraction is expected to  increase from $\sim$0.15 to 0.4 upon adding water. We therefore infer that revived shallow NV$^-$ centres, which exhibit reduced $T_1$  times, contribute more strongly to the ensemble-averaged signal in water, leading to the observed decrease in coherence time. 
This interpretation is further supported by the confocal PL spectra measurements in Fig.~\ref{fig:ND_H20}(c), which reveal a pronounced enhancement of NV$^-$ emission upon addition of water by about 50\% at close to the zero-phonon line (ZPL) at 637~nm. %Given ambient water adsorption \cite{li2026nanoscale} (reflected in the dashed lines in Fig.~\ref{fig:ND_H20}(b)), as well as deviations from the ideal spherical geometry, and the fact that the bulk NV$^0$ centers are not included in the simulations, the model and experimental results show reasonable agreement.

Finally, all three nanodiamond sizes were exposed to 5~$\mu$L of an aqueous Mn$^{2+}$ solution (500~$\mu$M), and all exhibit a strong reduction of $T_1$ (Fig.~\ref{fig:ND_H20}(d)). For ND100, $T_1$ decreases by a factor of $\sim2.5$. In turn, the smaller ND50 particles show much higher sensitivity, decreasing by a factor of $\sim$6.3. For comparison, the sensitivity to Gd$^{3+}$ ($S$ = 7/2) at the same concentration in state-of-the-art quantum relaxometry probes is a factor of 4 \cite{perona2020nanodiamond}. However, as shown above, the high sensitivity of ND50 comes at a cost: it is strongly influenced by environmental fluctuations, such as variations in water content, which can complicate calibration. By contrast, ND70 offers a favorable compromise, maintaining high sensitivity ($T_1$ changes by a factor of $\sim5$) while being more robust to non-target environmental effects. Moreover, for ND70, measurements with a high SNR were completed within 2~min (Supplementary Fig.~6), in stark contrast to the $\sim1440$~min required for conventional time-resolved protocol in Fig.~\ref{fig:bulk}(e). \REVISION{Even when compared to optimized protocols reported in the literature, this measurement remains among the fastest achieved, corresponding to a speedup of $\sim$2.5--30 times (see Supplementary Table~1). Note that we also investigated possible heating effects arising from microwave irradiation of this power and found no measurable change in local temperature (see Supplementary Note~3). It is also worth noting that the optical load imposed on the sample by our scheme is comparatively low. Established nanodiamond relaxometry protocols typically quote laser powers in the microwatt range, but this power is delivered through a diffraction-limited focus, so that the resulting intensity at the sample reaches $\sim$$10^5$~W/cm$^2$ \cite{perona2020nanodiamond}. In our wide-field configuration the same or higher total power is distributed over a millimetre-scale spot, giving 0.04--1.1~W/cm$^2$, i.e.\ five orders of magnitude lower (Supplementary Table~1). Since photobleaching and light-induced damage scale with intensity and dose rather than with total power, the optical perturbation of a biological sample in our geometry is substantially weaker than in confocal excitation, and the microwave field adds no measurable thermal load on top of it.} Moreover, in applications outside biology, where fluctuations in water content are less critical, ND50 becomes the optimal choice. Its intrinsically high $T_1$ tolerates stronger laser and MW excitation (Fig.~\ref{fig:ND_MW_LP}), enabling high-accuracy measurements with acquisition times potentially reduced to only a few seconds.

%\section*{Conclusions}

In conclusion, we have introduced a parameter-free FDR protocol for determining the ground-state $T_1$ time of a spin qubit under continuous wave illumination, establishing a direct bridge to conventional pulsed measurements. By eliminating fitting ambiguities, the method provides a robust and quantitative route to relaxation dynamics in realistic experimental environments. Our results elucidate the pronounced influence of water on $T_1$ in small nanodiamonds and demonstrate high sensitivity to micromolar concentrations of high-spin ions. The frequency-resolved implementation is fully compatible with confocal microscopy, provided laser-induced contributions are properly accounted for. Operating optimally for $T_1$ values up to $\sim$200~ms, where low-frequency drift and $1/f$ noise remain negligible, our FDR approach encompasses essentially all known nanoscale spin qubits and can in principle be used not only for ODMR but also for other types of magnetic resonance methods (PDMR, ESR, etc.). %Beyond offering a practical metrological tool, this framework enables fast probing of spin relaxation processes, paving the way for advanced high-throughput sensing applications at nanoscale.
\REVISION{Currently, the approach has been demonstrated on ensembles of NV centers, and it will be interesting to optimize it for confocal measurements on individual nanoparticles or photonic nanostructures, where lock-in detection of signals from single NV centers has already been demonstrated \cite{pershin2025coherence}. Importantly, this framework provides a practical metrological tool for reliably characterizing long relaxation times, which are otherwise difficult to access with conventional pulsed schemes within practical measurement times. Such capability is directly relevant to quantum computing, where long and well-characterized relaxation times are critical for qubit performance and error correction, and to quantum communication, where robust quantum memories and repeaters rely on long $T_1$ times}. Taken together, the FDR method may lead to a real breakthrough in quantum relaxometry, providing an accurate, rapid, sensitive, and experimentally efficient route to measurements in biologically relevant environments and beyond.

\section*{Methods}\label{methods}

\subsection*{Samples preparation}

To study longitudinal relaxation in ensembles of negatively charged NV centers, we used two $(100)$ single-crystal diamond bulk samples grown by chemical vapor deposition (CVD) method: sample DNV-B1 from Element Six UK Ltd. \cite{e6cvd_homepage} ([N] = 800~ppb, \REVISION{$5\times5\times0.5$~mm$^3$} "DNV") and a single-crystal sample from Diamond Elements Pvt. Ltd. \cite{diamond_elements_homepage} ([N] = 10~ppm, \REVISION{$3\times3\times0.3$~mm$^3$} "IN3x3"). We also used high-pressure high-temperature (HPHT) nanodiamonds of different sizes doped with NV$^-$ centers: 50~nm ("ND50"), 70~nm ("ND70"), 100~nm ("ND100"), as well as \REVISION{140~nm (for investigating the heating effects only)} purchased from Adamas Nanotechnologies Inc. \cite{adamasnano_website}, to prepare samples by depositing $\sim$5~$\mu$g of material on a 100~$\mu$m thick non-luminescent borosilicate glass plate using the drop-casting method. Mn$^{2+}$ solutions were prepared by dissolving manganese nitride (99.995\%, ThermoScientific) in ultrapure water. The liquid samples were sandwiched between two glass plates to prevent solvent evaporation during the measurements.

\subsection*{cw-ODMR setup and FDR measurement protocol}

cw-ODMR measurements were performed using a home-built optical setup based on the Newport integrating sphere 819C-IS-5.3 (IS), which contains four connection ports. The principal scheme of the setup is shown in Fig.~\ref{fig:1_FRS_IN}(c). The sample was placed on a coplanar waveguide antenna and connected to the bottom port of the IS. A 520~nm fiber coupled laser from Roithner, delivering up to 300~mW of power to about 6~mm diameter spot through a 550~nm dichroic mirror (Thorlabs DMLP550T), served as an excitation source and was connected to a side port of the IS. The radiation from the sample was collected through a biased silicon-based free-space photodetector (Thorlabs DET100A2) using a 700~nm long-pass hard coated filter (ThorLabs FELH0700) and 645~nm long-pass colored glass filter (Thorlabs FGL645M) and connected to the top port of the IS. A colored glass filter was used to prevent scattered laser beams from the inner surface of the sphere from reaching the photodetector input. The photocurrent from the detector output was converted into voltage using a 1~MHz bandwidth transimpedance amplifier (TIA) (Femto DHPCA-100) and connected to the input of a lock-In amplifier (Anfatec USBLockIn250).

The microwave (MW) field near the ground state spin resonance frequency (2870 $\pm$ 70~MHz) was generated by a MW generator (Vaunix LabBrick LSG-402) in conjunction with a high-power amplifier (Mini-Circuits ZHL-25W-63+) and applied to the sample through a coplanar waveguide \REVISION{(2~mm width)} terminated with a high-power (100~W) 50~$\Omega$ termination. The microwave excitation amplitude was modulated using a high-speed switch (Mini-Circuits ZASWA-2-50DRA+) and controlled by a TTL signal from the internal reference source of the lock-in amplifier. The modulation frequency $f_{\mathrm{m}}=\omega_{\mathrm{m}}/2\pi$ varied in the range from 1~Hz to 20~kHz, and the duty cycle was set equal to $D = 0.5$, which corresponds to symmetric on/off modulation. This is not a small modulation amplitude case, but the first harmonic amplitude of the modulation can be treated as an effective $\delta \Omega$ in equation (\ref{eq3}), the functional form of $H(\omega_{\mathrm{m}})$ remains the same, and the lock-in amplifier isolates this harmonic at the modulation frequency $\omega_{\mathrm{m}}$. The microwave power on the sample was calibrated and could be varied over a wide range to study linear and saturated cw-ODMR response modes. 

A setup for the time-domain measurements was described in \cite{pershin2025coherence}. The following protocol was used for the TDR measurements (see inset in Fig.~\ref{fig:bulk})(e)). The NV centers are initially polarized into the $|m_s = 0\rangle$ state by an optical pulse lasting 1~ms. Next, a microwave $\pi$-pulse are applied, typically for 250~ns. Next, after the free evolution time $\tau_{\mathrm{del}}$, a second optical pulse is applied for the readout ($R$) lasting 5~$\mu$s. The system is then allowed to fully relax over time $\tau_{\mathrm{res}}$. After this, the same sequence is applied, but with the $\pi$-pulse omitted, and the resulting time  $T_1$ is derived from the exponential decay of the $1-R(\pi_\text{on})/R(\pi_\text{off})$ function. The cryogenic part of this setup was also used for the low temperature measurements without using the objective. Photoluminescence (PL) spectra were recorded using a Renishaw inVia Raman Microscope with a 50$\times$ Leica objective. All signals were recorded in the Earth's magnetic field.

\section*{Contributions}

V.V.\ and O.A.\ developed the research methodology. V.V., A.P., and A.G.\ conceived the work. V.V., Ch.B., and A.P.\ carried out the experiments. All authors discussed the results. V.V.\ and A.P.\ wrote the manuscript with the contribution of all authors. A.G.\ secured the funding and supervised his group members together with A.P.

%\section{Corresponding authors}

%Correspondence to Vladimir Verkhovlyuk (vladimir.verkhovlyuk@wigner.hun-ren.hu), Anton Pershin (pershin.anton@wigner.hun-ren.hu) or Adam Gali (gali.adam@wigner.hun-ren.hu).

\section*{Competing interests}

The authors declare that there are no competing interests.

\section*{Acknowledgments}

The authors thank D.\ Beke and Sz.\ Czene for
experimental assistance and fruitful discussions. A.G.\ acknowledges the support from European Commission within Horizon Europe projects QuSPARC and SPINUS (Grant Nos.\ 101186889 and 101135699). A.P.\ acknowledges the financial support of J\'anos Bolyai Research Fellowship of the Hungarian Academy of Sciences. Ch.B.\ is grateful for support from the Stipendium Hungaricum scholarship.

%%\bibliography{main}


\clearpage

%%%%%%%%%%%%%%%%%%%%%%%%%%%%%%%%%%%%%%%%%%%%%%%%%%%%%%%%%%%%%%%%%%%%%%%%%%%%%
%%  SUPPLEMENTARY INFORMATION
%%  (separate document in the journal submission; appended here so that the
%%   arXiv version shows the main text and the SI together)
%%%%%%%%%%%%%%%%%%%%%%%%%%%%%%%%%%%%%%%%%%%%%%%%%%%%%%%%%%%%%%%%%%%%%%%%%%%%%
\setcounter{figure}{0}
\setcounter{table}{0}
\setcounter{section}{0}
\setcounter{equation}{0}
\renewcommand{\figurename}{Supplementary Figure}
\renewcommand{\tablename}{Supplementary Table}
\renewcommand{\thefigure}{\arabic{figure}}
\renewcommand{\thetable}{\arabic{table}}
\renewcommand{\theequation}{S\arabic{equation}}
\renewcommand{\thesection}{Supplementary Note~\arabic{section}}

\begin{center}
{\LARGE\bfseries Supplementary Information}\\[0.6em]
{\large Quantum Relaxometry Under Continuous Wave Excitation}
\end{center}
% keep the SI figures from floating above the SI title
\FloatBarrier

\vspace{1em}

\section{Bulk samples}

\begin{figure}
    \centering
    \includegraphics[width=0.75\linewidth]{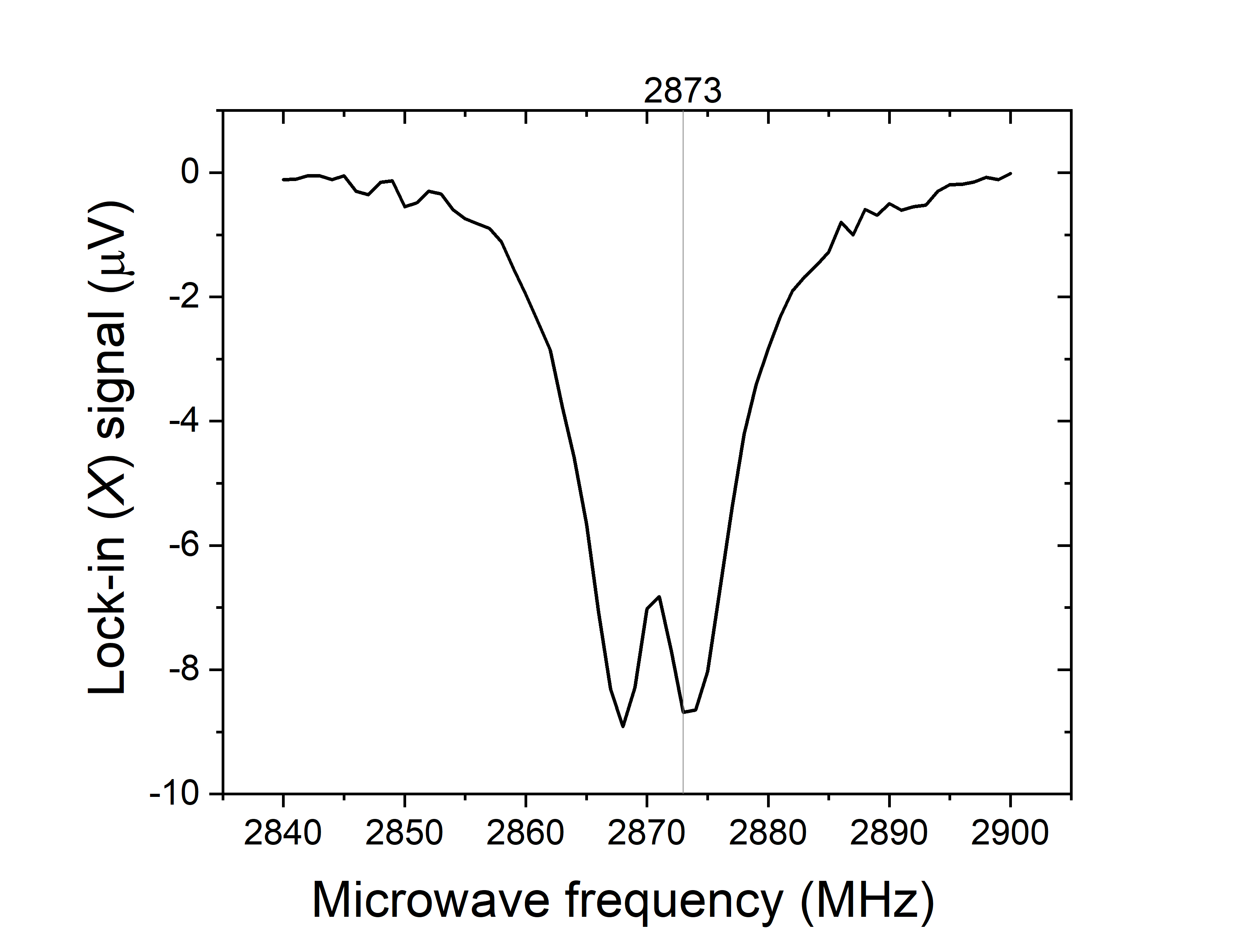}
    \caption{cw-ODMR spectum recorded in the DNV sample at 20~dBm microwave power and 0.3~W/cm$^2$ laser intensity.}
    \label{smfig:DNV_ODMR}
\end{figure}
%Supplementary Fig.~\ref{smfig:DNV_ODMR} shows a cw-ODMR spectrum for the DNV sample recorded under continues wave microwave and laser excitation using a lock-in amplifier. This is the first step of the FDR measurement protocol, followed by fixing the microwave frequency near the maximum contrast (2873~MHz in this specific case) and measuring the lock-in amplifier signals $X$ and $Y$ while sweeping the modulation frequency. For the TDR protocol a $\pi$-pulse of the same microwave frequency at maximum contrast was used. To determine the optimal pulse duration Rabi measurements were performed. Results for both bulk samples are shown in Supplementary Fig.~\ref{smfig:Rabi} as well as measurement protocol. NV$^-$ centers were initially polarized to the $|m_s = 0 \rangle$ state with a 4~$\mu$s optical pulse. A microwave pulse of variable duration $\tau$ was then applied, followed by a second optical pulse. To readout the spin state, two 500~ns pulses were applied at the beginning (signal) and end (reference pulse) of the second 4~$\mu$s optical pulse. As a result, we obtain photoluminescence oscillations at a Rabi frequency $\Omega$, damped with a certain coherence time $T_2^{Rabi}$. These parameters can be extract from the simulation using the expression 

In this Supplementary Note, we have collected the supporting experimental data for bulk samples. Supplementary Fig.~\ref{smfig:DNV_ODMR} shows a cw-ODMR spectrum of the DNV sample recorded under continuous microwave and laser excitation using lock-in detection. This measurement constitutes the first step of the FDR protocol. The microwave frequency is subsequently fixed near the point of maximum ODMR contrast (2873~MHz in this case), while the lock-in amplifier signals $X$ and $Y$ are recorded as a function of the modulation frequency.

For the TDR protocol, a $\pi$-pulse at the same microwave frequency corresponding to the maximum ODMR contrast was employed. The optimal pulse duration was determined from Rabi measurements. The results for both bulk samples, together with the measurement protocol, are shown in Supplementary Fig.~\ref{smfig:Rabi}. NV$^-$ centres were first polarised into the $|m_s = 0 \rangle$ state using a 4~$\mu$s optical pulse. A microwave pulse of variable duration $\tau$ was then applied, followed by a second optical pulse. The spin state was read out using two 500~ns detection windows placed at the beginning (signal) and end (reference) of the second 4~$\mu$s optical pulse.

This protocol yields photoluminescence oscillations at the Rabi frequency $\Omega$, damped with a coherence time $T_2^{\mathrm{Rabi}}$. These parameters are obtained by fitting the signal with

$$S(t)=A+B\cos(\Omega \tau)e^{-\tau/T_2^{Rabi}}$$. 

%The simulated curves and parameters are shown in Supplementary Fig.~\ref{smfig:Rabi}. 
%The $\pi$-pulse duration is defined as $\tau(\pi)=1/(2\Omega)$. For DNV the $\pi$-pulse of 238~ns was found substantially shorter than $T_2^{Rabi}$ of 1$\mu$s (Supplementary Fig.~\ref{smfig:Rabi}(a)). In this case, up to five oscillations are observed, allowing for a well-defined $\pi$-pulse. However, for IN3x3, the $\pi$-pulse 357~ns is longer than $T_2^{Rabi}$ of 0.23$\mu$s (Supplementary Fig.~\ref{smfig:Rabi}(b)). Here, only a half of an oscillation is seen and the $\pi$-pulse can only defined only approximately. The precise resolution of the $\pi$-pulse for IN3x3 would require MW power well exceeding 50 dBM, which it technically very challenging.
The fitted curves and extracted parameters are shown in Supplementary Fig.~\ref{smfig:Rabi}. The $\pi$-pulse duration is defined as $\tau_\pi = 1/(2\Omega)$. For the DNV sample, the $\pi$-pulse duration of 238~ns is substantially shorter than the estimsted $T_2^{\mathrm{Rabi}}$ of 1~$\mu$s (Supplementary Fig.~\ref{smfig:Rabi}a). In this case, up to five Rabi oscillations are observed, allowing precise determination of the $\pi$-pulse length. In contrast, for the IN3x3 sample the $\pi$-pulse duration (357~ns) exceeds the measured $T_2^{\mathrm{Rabi}}$ of 0.23~$\mu$s (Supplementary Fig.~\ref{smfig:Rabi}b). As a result, only about a half of an oscillation is visible and the $\pi$-pulse duration can be determined only approximately. Achieving a more precise determination for IN3x3 would require microwave powers well exceeding 50~dBm, which is technically challenging.

\begin{figure}
    \centering
    \includegraphics[width=0.75\linewidth]{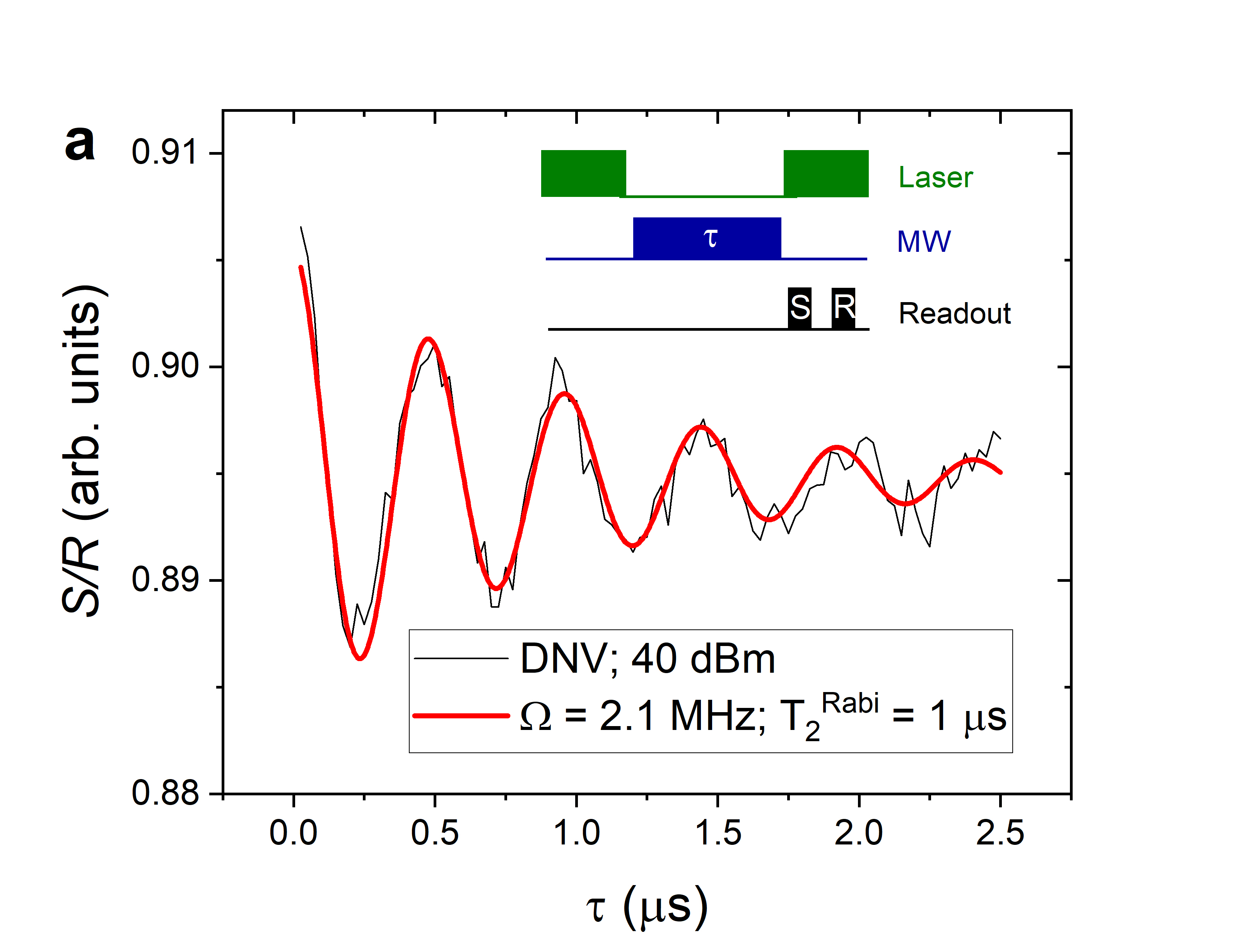}
    \includegraphics[width=0.75\linewidth]{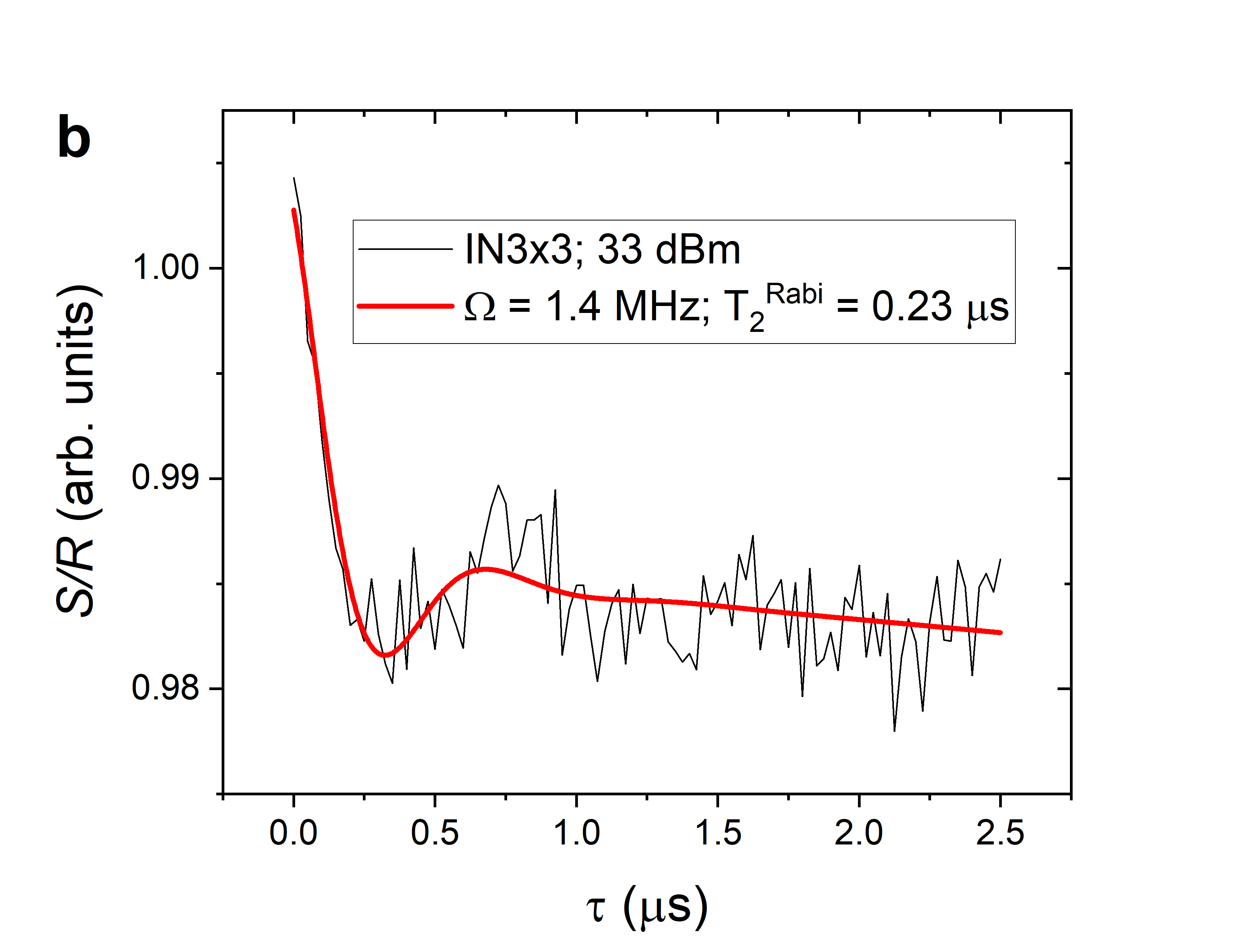}
    \caption{Rabi measurement results for DNV (a) and IN3x3 (b) samples. The inset in (a) shows the measurement protocol.}
    \label{smfig:Rabi}
\end{figure}

Results of low-temperature pulsed TDR measurements for the DNV sample are shown in Supplementary Fig.~\ref{smfig:low_conf_DNV}. As noted in the main text, even after 48 hours of data acquisition the signal-to-noise ratio remains very low, preventing a reliable determination of the $T_1$ time. Nevertheless, two fitting approaches were applied to quantify the signal. An unconstrained fit yields $T_1 = 468$ ms, with an uncertainty of 480 ms exceeding the fitted value itself. A more reasonable estimate is obtained from a constrained fit, taking into account that the signal approaches a plateau after $\sim$700 ms. In this case, the extracted relaxation time is $T_1 = 223$ ms with an uncertainty of 107 ms. These estimates are at least consistent in order of magnitude with the value of $\sim200$ ms obtained using the FDR protocol. 

\begin{figure}
    \centering
    \includegraphics[width=0.75\linewidth]{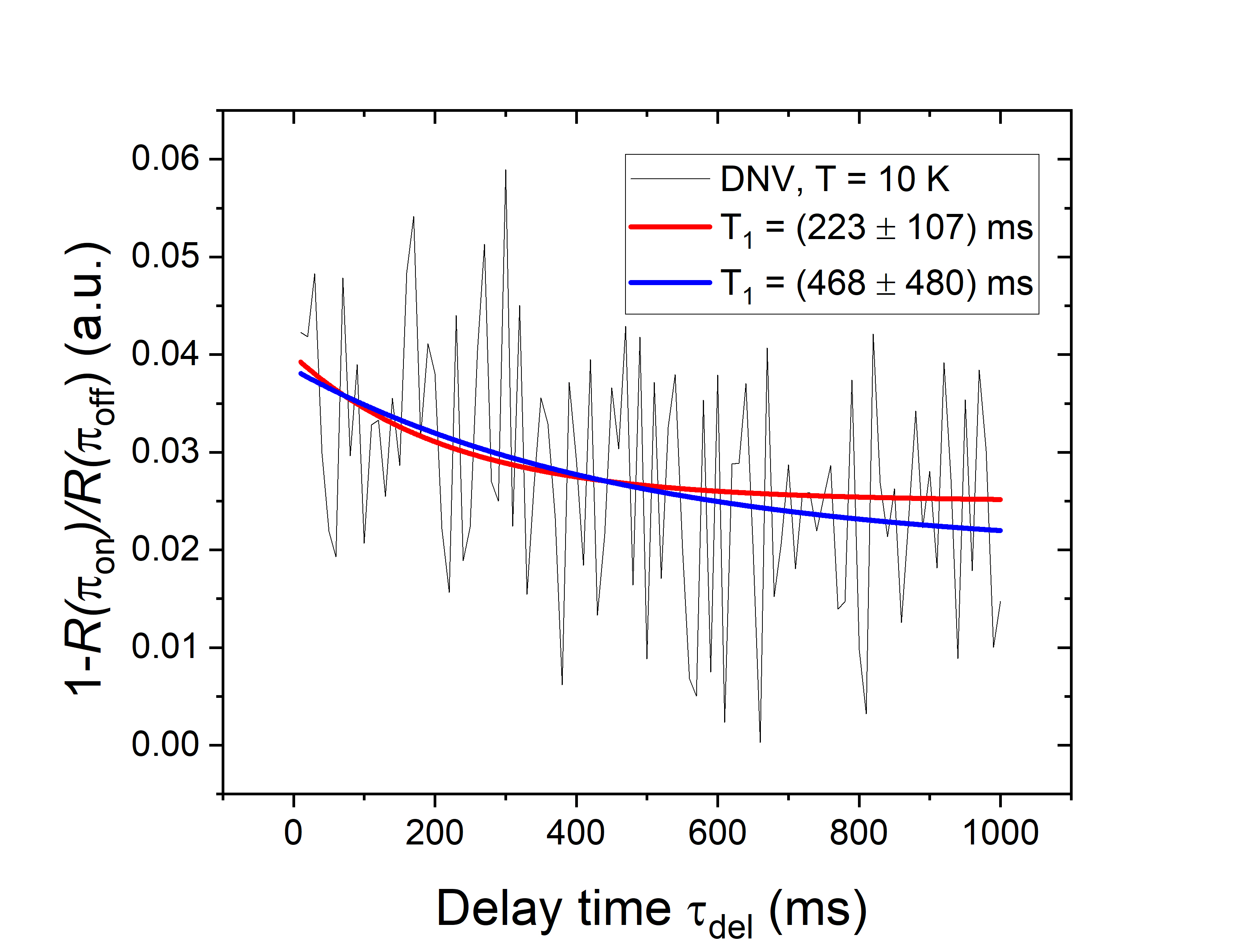}
    \caption{Results of pulsed TDR measurements at low temperature ($T=10$~K) in the DNV sample. Measurement time is 48 hours. Solid lines show two possible decaying exponential simulation options.}
    \label{smfig:low_conf_DNV}
\end{figure}

\REVISION{
For the DNV sample, the laser-intensity dependence of $1/T_{\mathrm{eff}}$ was
recorded at three microwave powers (Fig.~2(d) of the main text). The
corresponding zero-intensity intercepts are
$\Gamma_0 = (190.7\pm15.9)$~s$^{-1}$ at 5~dBm,
$(180.3\pm6.5)$~s$^{-1}$ at 15~dBm and
$(194.1\pm4.3)$~s$^{-1}$ at 25~dBm, i.e.\ $T_1 = 5.24$, 5.55 and 5.15~ms,
respectively. The three determinations agree with one another within their
uncertainties, which confirms that the extrapolation is insensitive to the
microwave power over this range, as expected in the linear-response regime.
The value quoted in the main text is the one obtained at 15~dBm. If instead a
single best estimate combining all three microwave powers is desired, their
weighted mean gives $\Gamma_0 = (189.9\pm4.4)$~s$^{-1}$, i.e.\
$T_1 = (5.26\pm0.12)$~ms, where the uncertainty has been scaled by the Birge
ratio $\sqrt{\chi^2/\nu} = 1.25$ ($\chi^2 = 3.14$ for $\nu = 2$ degrees of
freedom) to account for the residual scatter between the three values. This
averaged value agrees with the quoted one to within $1.2$ standard deviations,
so the choice does not affect any of the conclusions drawn here.
}

\REVISION{
\section{Microwave saturation effect}

This Supplementary Note presents data demonstrating the effect of microwave-induced spin saturation on the response signal of a lock-in amplifier ($Y$). Supplementary Fig.~\ref{smfig:SI_IN_mw}(a) shows the $Y$ components of the lock-in signal as a function of modulation frequency for the IN3x3 sample at various microwave powers, as well as simulations using equation
\begin{equation}
Y(f_{\mathrm{m}}) =  -\frac{2\pi f_{\mathrm{m}} T_{\mathrm{eff}}}{1 + (2\pi f_{\mathrm{m}} T_{\mathrm{eff}})^2}. \label{si_eq1}
\end{equation}
The figure shows that when moving from the linear approximation mode to the microwave saturation mode (see main Fig. 2(a)) the signal shape becomes less consistent with the equations (\ref{si_eq1}): in addition to the shift of the maximum to higher frequencies, an additional broadening contribution is visible, which increases with increasing microwave power. This must be taken into account in the analysis, However, the position of the signal maximum still correctly reflects the relaxation contribution and can be used in relative measurements. For the DNV sample (Supplementary Fig.~\ref{smfig:SI_IN_mw}(b)) we see a fairly good agreement between the experimental and fit data over the entire microwave power range, meaning again that for this sample we are always on a plateau. It is also worth noting that the conformity to equation~(\ref{si_eq1}) may indicate that we are in the linear approximation mode. 
\begin{figure}
    \centering
    \includegraphics[width=0.75\linewidth]{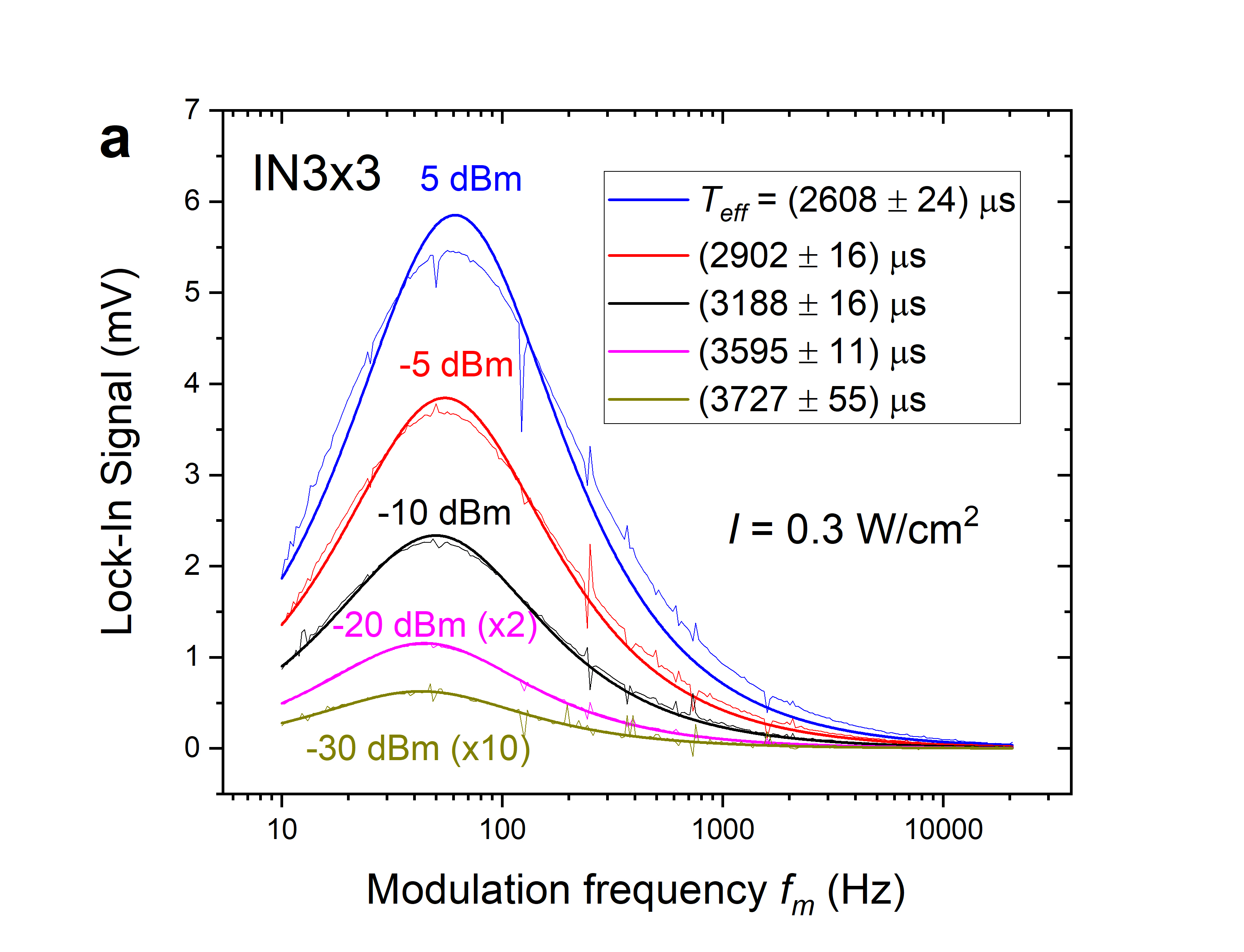}
    \includegraphics[width=0.75\linewidth]{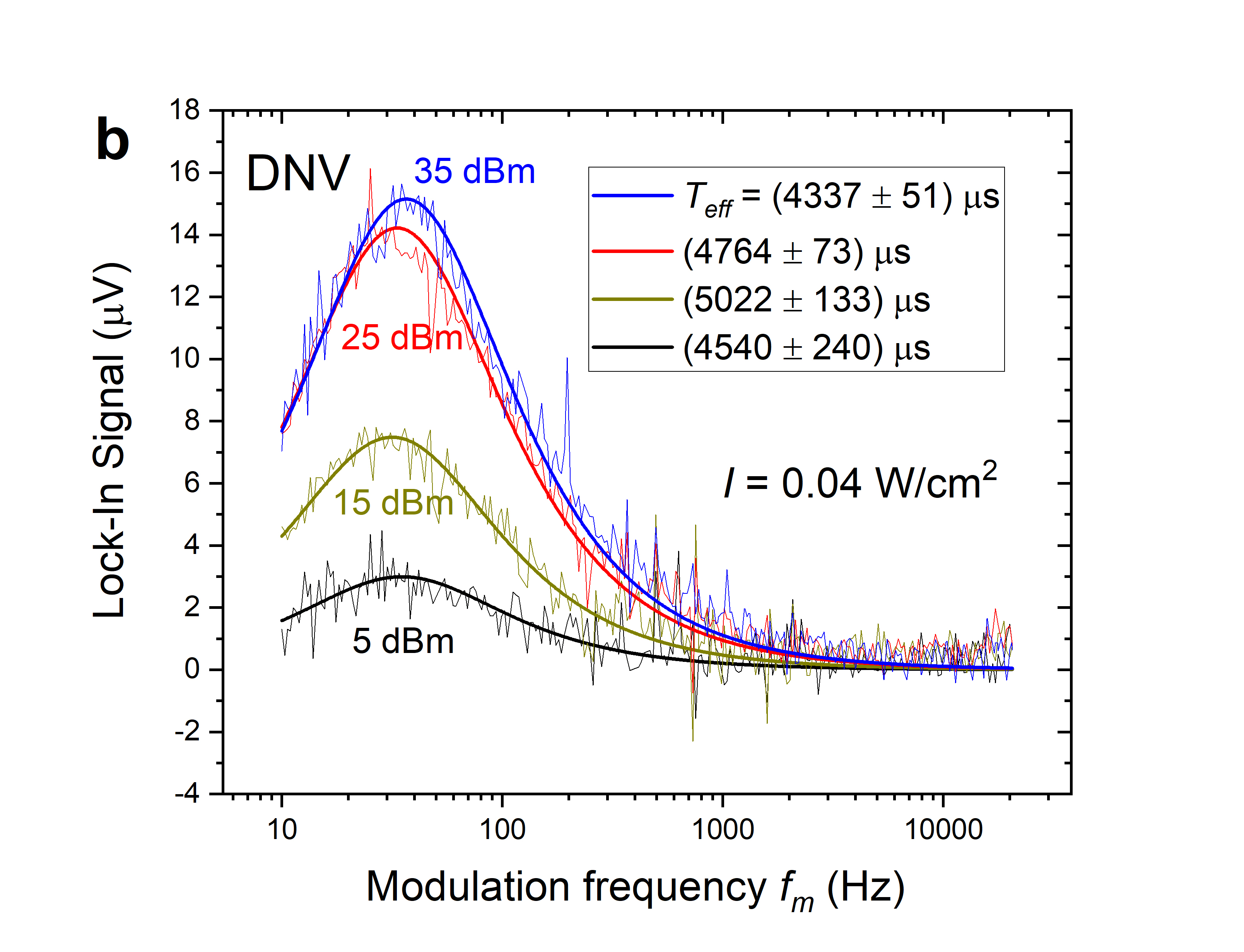}
    \caption{Dependence of the $Y$ signal on the microwave power for IN3x3 (a) and DNV (b) samples. Simulation data are shown as solid lines. $I$ is the laser intensity.}
    \label{smfig:SI_IN_mw}
\end{figure}
}

\section{Nanodiamond samples}
\begin{figure}
    \centering
    \includegraphics[width=0.75\linewidth]{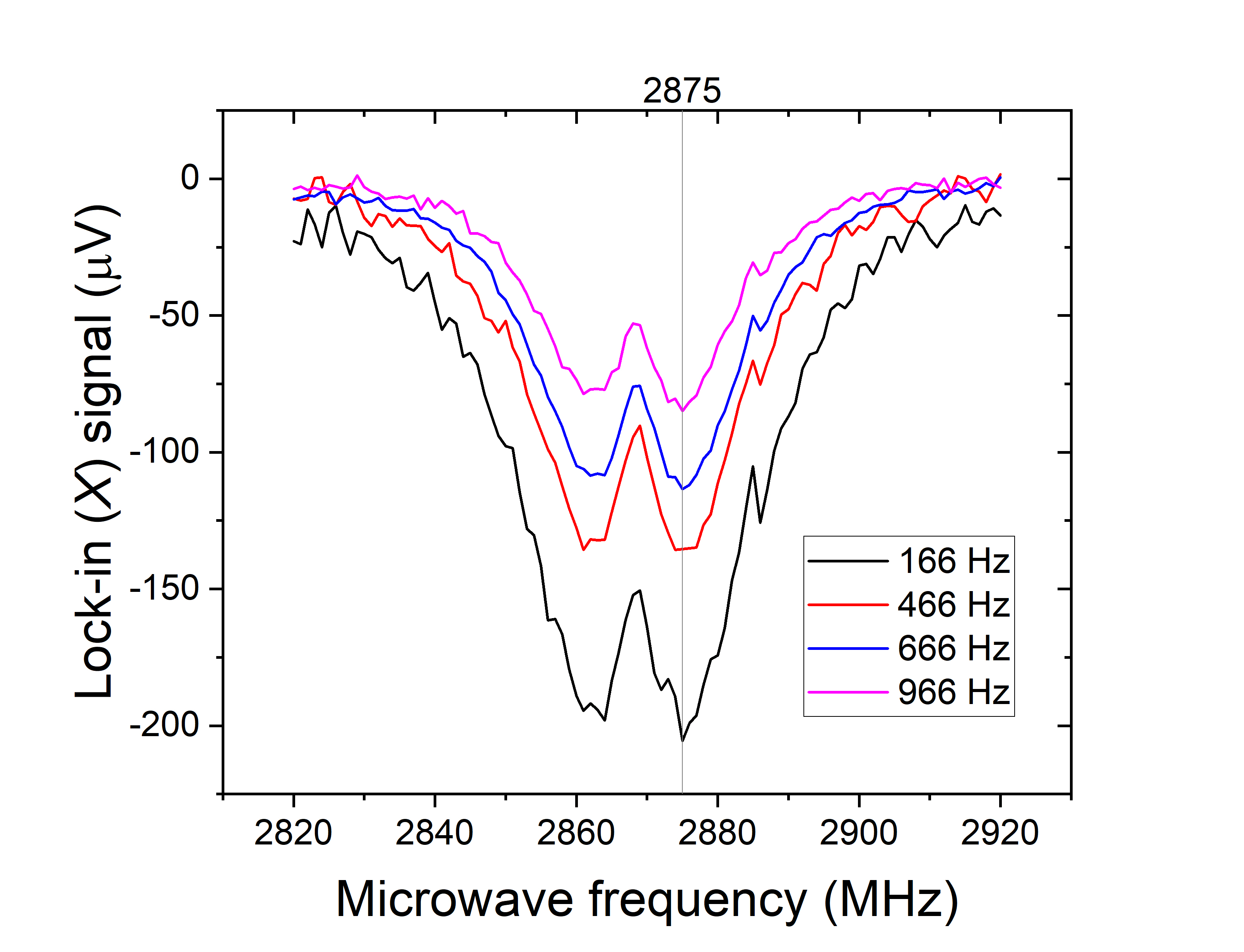}
    \caption{ODMR spectra of ND100, obtained at the specified values of modulation frequency.}
    \label{smfig:ND100_ODMR}
\end{figure}

In this Supplementary Note we present experimental data obtained from nanodiamond samples. Supplementary Fig.~\ref{smfig:ND100_ODMR} shows ODMR spectra of ND100 measured at different modulation frequencies. The lock-in contrast decreases across the entire spectrum as the modulation frequency increases. As discussed in the main text, this behavior corresponds to the response of the $X$ lock-in component and enables direct extraction of the effective relaxation time $T_{\mathrm{eff}}$ from the ODMR measurements.

\begin{figure}
    \centering
    \includegraphics[width=0.75\linewidth]{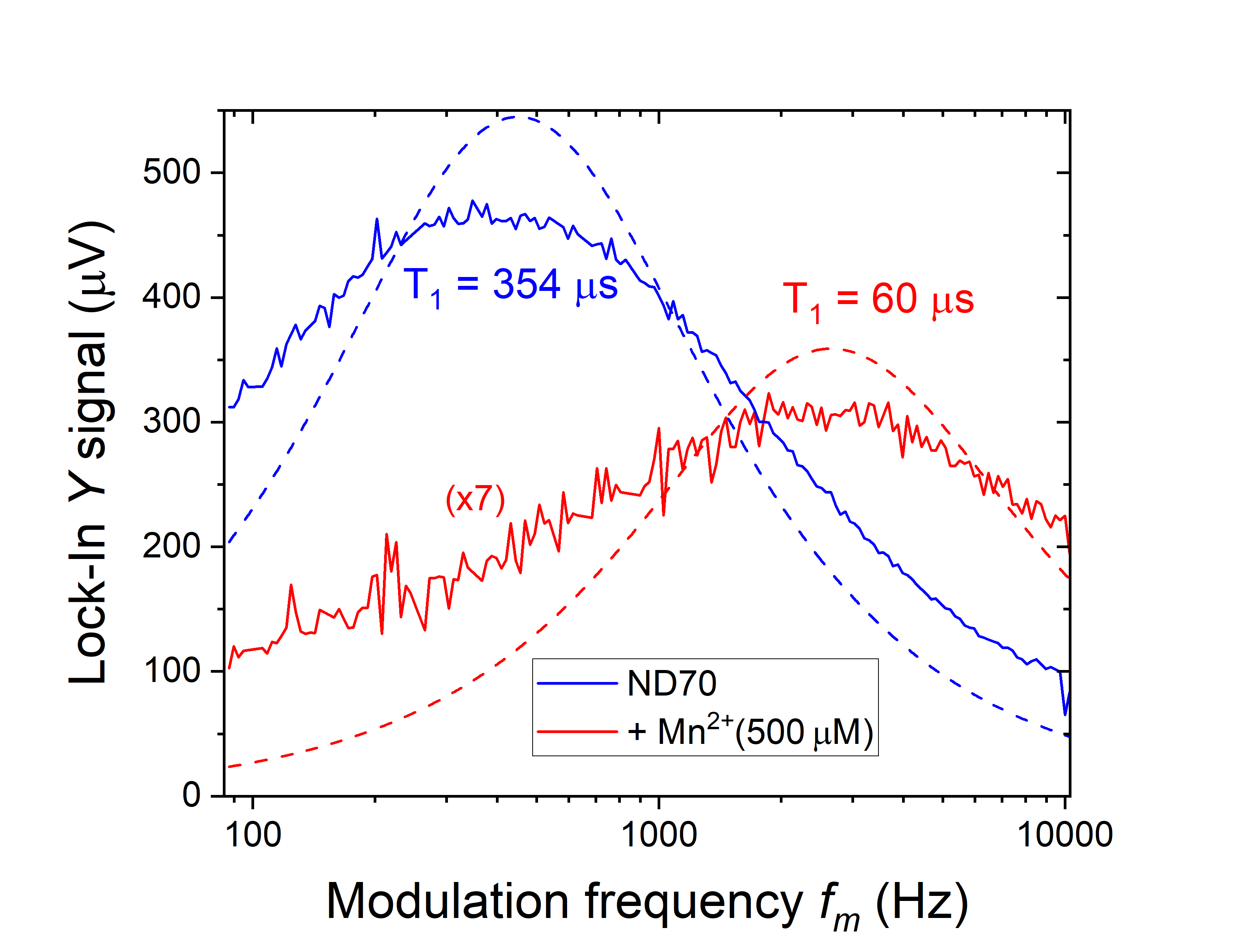}
    \caption{$Y$ component of lock-in signal for ND70 as a function of modulation frequency in a powder sample and in aqueous solution of Mn$^{2+}$. The best simulation data are shown as dashed lines.}
    \label{smfig:ND70_Mn}
\end{figure}

Supplementary Fig.~\ref{smfig:ND70_Mn} shows the typical behaviour of the $Y$ lock-in component for ND70 after the addition of Mn$^{2+}$ ions. The lock-in contrast decreases by about an order of magnitude, primarily due to a 5.5-fold reduction of the $T_1$ relaxation time. \REVISION{Therefore, we used +15~dBm MV power to increase the signal-to-noise ratio and to maintain the same power level across different measurements. Indeed, in this case, we are not in the linear approximation regime anymore (consistent with Figure 4) and the shape of signals deviates from the one set by equation~(\ref{si_eq1}). However, we used the positions of the signal maxima to determine the relative contribution of the spin-spin interaction from manganese ions to the spin-lattice relaxation. } The measurements were performed over 256 modulation frequencies using a lock-in time constant of 500~ms, resulting in a total acquisition time of $\sim$2 minutes. \REVISION{Relative comparison of this measurement time with the literature data, as well as other related parameters, such as laser intensity, are provided in Supplementary Table~\ref{tab:speed_comparison}.}

\begin{table}[htbp]
\centering
\footnotesize
\begin{tabular}{@{}p{3.4cm}lcc@{}}
\toprule
& laser at sample & intensity (W/cm$^2$) & acq.\ per $T_1$ (min) \\
\midrule
Ref.~\cite{perona2020nanodiamond}
  & 100~$\mu$W, confocal & $2\times10^{5}$ & $\approx$8 \\
Ref.~\cite{sigaeva2022diamond}, in cells
  & 31~$\mu$W, NA 1.0 & $2\times10^{5}$ & $\approx$10 \\
Ref.~\cite{sigaeva2022diamond}, in solution
  & 500~$\mu$W & $7.3\times10^{5}$ & --- \\
Ref.~\cite{wu2023diamond}
  & 50~$\mu$W, NA 1.40 & --- & $\approx$5 \\
Ref.~\cite{grant2023method}
  & 160~mW, 100~$\mu$m waist & $\approx2\times10^{3}$ & 30--60 \\[1mm]
\textbf{This work} & wide field, 6~mm spot
  & $\mathbf{0.04}$--$\mathbf{1.1}$ & $\mathbf{\approx2}$ \\
\bottomrule
\end{tabular}
\caption{Comparison of laser excitation conditions and acquisition times for $T_1$ relaxometry measurements reported in the literature versus this work.}
\label{tab:speed_comparison}
\end{table}

\begin{figure}
    \centering
    \includegraphics[width=0.75\linewidth]{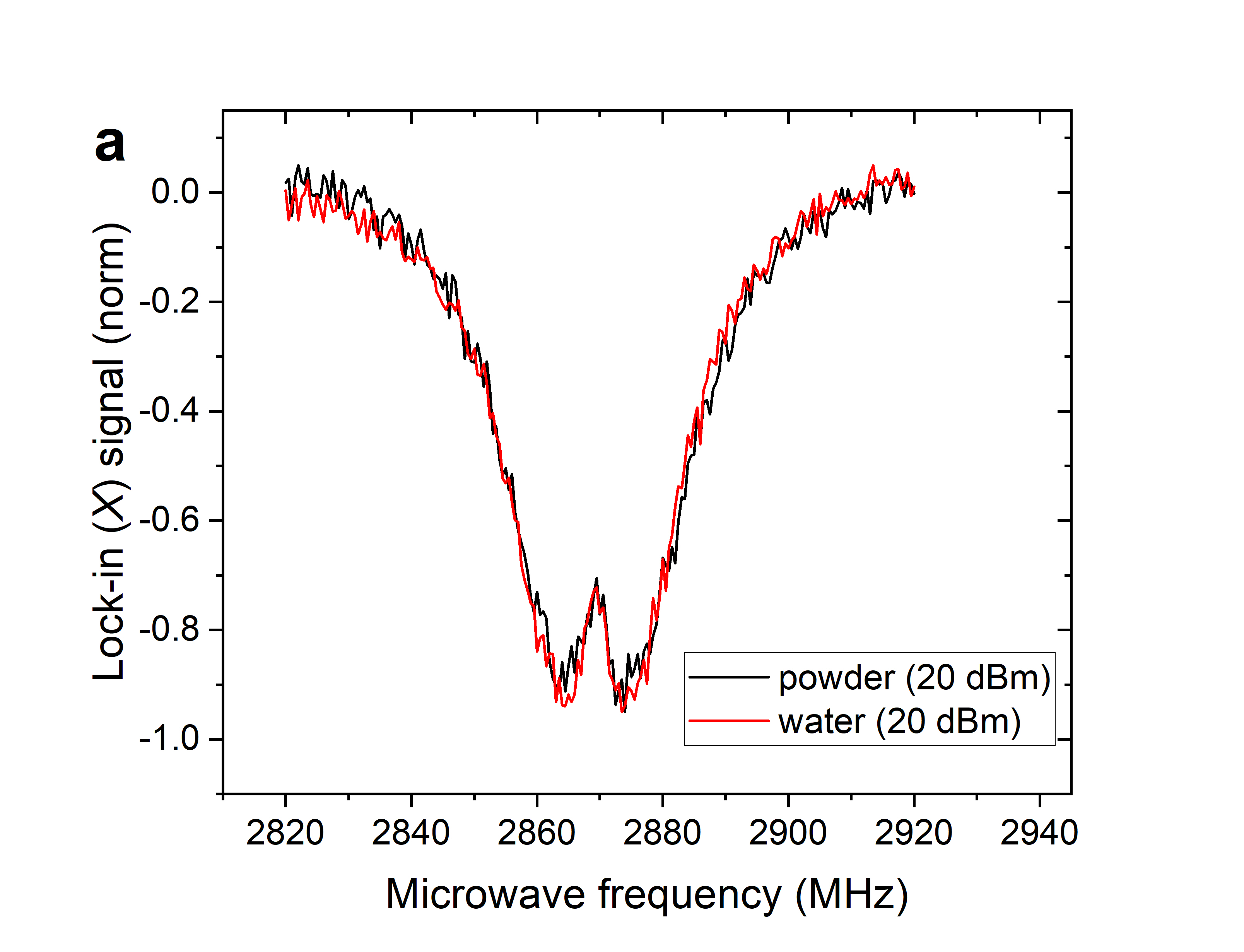}
    \includegraphics[width=0.75\linewidth]{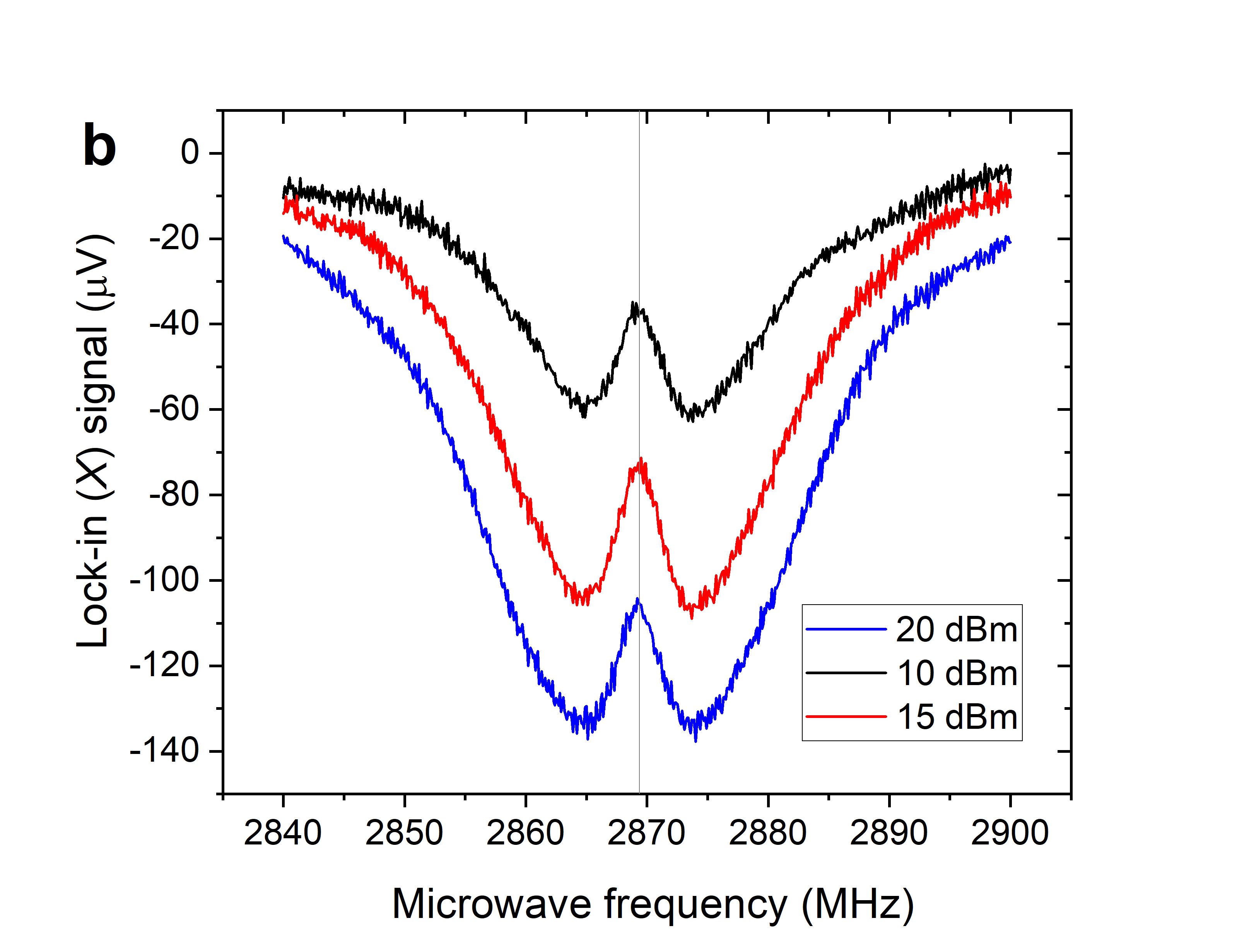}
    \caption{(a) ODMR spectra of ND140 sample in dry powder and in water, obtained at the same microwave power of 20 dBm. (b) ODMR spectra of ND140 sample in water, obtained at different microwave powers.}
    \label{smfig:ND140_ODMR}
\end{figure}
\REVISION{
For potential application of biosensors in living cells, it is crucial to avoid additional heating of the sample with microwave power, especially in aqueous solution, where the permittivity is quite high. % ($\varepsilon = 76.7$, $\varepsilon = 10.6$,
%$\sigma_{\mathrm{eff}} = 1.70$ S/m at 25$^\circ$C, from the Liebe
%\textit{et al.} double-Debye parameterization). 
An increase in sample temperature upon application of microwave radiation could manifest itself in the cw-ODMR spectra as a shift of the spectrum center, corresponding to the zero-field splitting constant $D$, to the low-frequency region with $\mathrm{d}D/\mathrm{d}T\approx-74$~kHz/K at room temperature (Acosta
\textit{et al.}, Phys.\ Rev.\ Lett.\ \textbf{104}, 070801 (2010))\cite{acosta2010temperature}. Supplementary Fig.~\ref{smfig:ND140_ODMR} shows the ODMR spectra of ND140, measured in powder and water at actual microwave powers, from which no temperature-induced changes in the constant $D$ is visible. Taking into account the spectral resolution of 100 kHz, we can conclude that the temperature of the samples was stable within $\sim$1~K. 
}

% Self-contained bibliography: the main text also carries its references inline
% as \bibitem entries, so no .bib file / BibTeX pass is needed for either document.

\end{document}